\documentclass[a4paper,USenglish,cleveref,autoref,thm-restate]{lipics-v2021}

\hideLIPIcs

\usepackage{pifont}% http://ctan.org/pkg/pifont
\newif\ifEditMode

\EditModetrue
\usepackage{preamble}
\usepackage{aliases}

\title{Airdrops: Giving Money Away Is Harder Than It Seems}

\author{Johnnatan Messias}{MPI-SWS
\and \url{https://johnnatan-messias.github.io}}{}{https://orcid.org/0000-0002-6021-8402}{}

\author{Aviv Yaish}{Yale University \and \url{https://www.avivyaish.com}}{}{https://orcid.org/0000-0002-7971-2494}{}

\author{Benjamin Livshits}{Imperial College London \and \url{https://www.doc.ic.ac.uk/~livshits}}{}{https://orcid.org/0000-0002-4921-8452}{}

\authorrunning{J. Messias, A. Yaish, and B. Livshits} 

\keywords{Airdrop design, Airdrop farmer, Web3, Layer 2} 

\ccsdesc[100]{ }

\acknowledgements{
J. Messias and A. Yaish contributed equally to this work. Some of this work was performed while the authors were at Matter Labs. \red{Version last revised on \today}.

}

\nolinenumbers %uncomment to disable line numbering

\begin{document}

\maketitle

\begin{abstract}
Airdrops are a popular mechanism used by blockchain protocols to bootstrap communities, reward early adopters, and decentralize token distribution. Despite their widespread adoption, the effectiveness of airdrops in achieving long-term user engagement and ecosystem growth remains poorly understood. In this paper, we present the first comprehensive empirical study of nine major airdrops across Ethereum and Layer-2 ecosystems. Our analysis reveals that a substantial share of tokens—up to 66\% in some cases—are rapidly sold, often in recipients' first post-claim transaction. We show that this behavior is largely driven by ``airdrop farmers,'' who strategically optimize eligibility criteria to extract value without contributing meaningfully to the ecosystem. We complement our quantitative findings with a case study of the Arbitrum airdrop, illustrating how short-term activity spikes fail to translate into sustained user involvement. Based on these results, we discuss common design pitfalls—such as Sybil vulnerability, poor incentive alignment, and governance token misuse—and propose actionable guidelines for designing more effective airdrop strategies.

\end{abstract}

%------------------------------------------------------------------------------
\section{Introduction} 
\label{sec:introduction}

Blockchain protocols often design reward programs to attract new customers and foster the loyalty of existing users.
In recent years, the practice of distributing platform-minted tokens, known as ``airdropping,'' has gained popularity \cite{makridis2023rise}.
For example, in 2023 alone,~4.56 billion US dollars worth of tokens were airdropped by various protocols~\cite{Report@Coingecko}.

Although the use of airdrops is prevalent in the blockchain world, we present preliminary findings showing that there is no significant correlation between performing an airdrop and a platform's popularity relative to available alternatives.
Intuitively, this is not optimal and may result in the loss of funds that can be used to improve the \gls{QoS} offered by the platform.

Although the fundamental concept of airdrops is relatively straightforward, the design space for such reward schemes is expansive and may vary depending on the characteristics of the platform implementing the airdrop. For instance, certain airdrop mechanisms focus on power users, granting them a substantial amount of rewards with the expectation that these users will stimulate valuable economic activity, subsequently attracting additional users.
However, a potential drawback to this approach arises when the distributed tokens empower users to propose changes to a protocol through decentralized governance, typically employing a one-token-one-vote strategy, where a single user may possess more than one voting token.
This introduces the risk of voting power concentration, where a minority of users hold the majority of the decision-making power~\cite{feichtinger2023hidden,messias2023understanding,sharma2024unpacking}.

To understand why previous airdrops did not always achieve their goals and to quantify their success, we first articulate a set of reasonable desired outcomes for airdrops.
We go over previous airdrops, see how they measure up, and uncover some interesting insights in the process, comparing them to a basic desiderata.
To ground our analysis in real-world observations, we focus on nine widely recognized airdrops: 1inch~\cite{bukov2020mooniswap}, Arbitrum~\cite{Kalodner-Usenix}, Arkham~\cite{arkham}, DYDX~\cite{juliano2018dydx}, ENS~\cite{ENS}, Lido~\cite{lido}, Optimism~\cite{optimism}, Tornado Cash~\cite{tornadocash}, and Uniswap~\cite{adams2021uniswap}.
These protocols were selected to ensure diversity across blockchain ecosystems (including Ethereum and Layer-2 rollups), protocol categories (e.g., DEXs, governance platforms, privacy tools, staking services), and market capitalizations (refer to Table~\ref{tab:airdrop-table}).
This curated selection allows us to draw generalizable insights while maintaining analytical depth across a representative cross-section of the blockchain landscape.

We conclude that most of the proceeds (as high as~65.75\%) have been sold via exchanges shortly after the airdrop, implying that the airdrops fall short of their intended purpose and benefit mostly \emph{airdrop farmers}, highly skilled users who employ sophisticated tactics to increase the share of tokens that they receive. 
One notable example of such farming activities occurred during the ZKsync Era. In this incident, a Sybil farmer orchestrated a fake airdrop, utilizing over \num{20000} accounts and deploying a fake, non-open-sourced DEX to qualify for the ZKsync token airdrop~\cite{Gemstone,Gemstone-Twitter}.
We analyze this incident and discuss it in~\S\ref{sec:gemstone}.
Furthermore, we describe the common challenges faced by airdrops in the past.
Given that airdrops are a relatively recent phenomenon, it is expected that the theoretical and practical understanding of them are still in their infancy, and thus previous airdrops may not have been entirely successful in the long run.
Finally, we use our analysis to suggest ways to create better airdrop mechanisms that are fairer to honest users.

\subsection{Research Questions}\label{sec:research-questions}

We address three critical research questions central to our study.

\paraibq{RQ-1: Do airdrops lead to long-term user retention?}

Airdrops are widely used to bootstrap user bases by distributing tokens to early adopters. However, their long-term effectiveness remains questionable. We examine whether protocols that conducted airdrops successfully retained users beyond the initial distribution period.

Our analysis shows that while airdrops temporarily increase user activity, they often fail to ensure lasting engagement. For instance, even widely publicized airdrops such as those conducted by Arbitrum or Optimism experienced an initial surge in user participation followed by a noticeable decline. These patterns are consistent with external reports, such as those by Messari~\cite{Bautista@Messari}, which observe that users often migrate to protocols offering new or better incentives once the airdrop rewards are exhausted.

Interestingly, user migration is not limited to protocols that conduct airdrops. Platforms without current airdrops also attract users motivated by the expectation of future distributions. Our findings suggest that airdrops may be necessary to attract users initially, but insufficient on their own to build a committed, long-term community.

\paraibq{RQ-2: What alternative incentives can enhance airdrops to sustain user engagement?}

Given the limitations of airdrops alone, we explore alternative mechanisms that could complement or extend their effectiveness. One promising strategy is to incentivize developers and builders to create applications such as \glspl{DEX}, \gls{NFT} platforms, and other dApps that enrich the ecosystem and enhance platform stickiness. This aligns user incentives with broader platform utility.

Protocols could also offer fee discounts or service credits to users holding airdropped tokens, creating ongoing utility for the asset beyond speculative value. Another viable approach is implementing multi-phase airdrops, allowing protocols to refine their criteria and distribution logic across rounds~\cite{allenCryptoAirdropsEvolutionary2023}. Finally, user-centric design such as incorporating community feedback into airdrop mechanisms can help protocols better align with user expectations and values.

However, these strategies must contend with user behavior patterns shaped by opportunism. Blockchain users are highly mobile and prone to chasing short-term gains. Without a compelling long-term value proposition, users are likely to disengage once rewards diminish.

\paraibq{RQ-3: What are the key design challenges faced when implementing airdrops?}

Airdrop campaigns face several technical and economic challenges that can undermine their objectives.
First, Sybil attacks remain a pervasive threat.
Sophisticated airdrop farmers create pseudonymous wallets to game eligibility criteria and maximize token claims~\cite{fanAltruisticProfitorientedMaking2023,Hu@WWW23,iqbal2021exploring,liu2022fighting,moser2022resurrecting,OTTE2020770,victor2020address}.
These actors may employ automated tools~\cite{Combine,NFTCopilot,FarmerFriends,SybilSamurai} to mimic legitimate user behavior and evade Sybil-detection heuristics.
Existing detection methods leverage graph analysis and machine learning~\cite{Zhou-ARTEMIS@WWW24,liu2022fighting,Celestia-airdrop}, but the adversarial nature of these users makes defense an ongoing arms race.

Second, governance token distributions pose their own risks.
When poorly designed, they may concentrate power in the hands of opportunistic or uninformed actors. Given that many governance systems rely on a one-token-one-vote mechanism, protocols risk empowering misaligned stakeholders~\cite{feichtinger2023hidden,messias2023understanding,sharma2024unpacking}. This undermines governance integrity and can discourage community participation.

Finally, other issues, such as insider trading before an airdrop, preferential access to airdrop distribution criteria or leaks related to forthcoming airdrops can erode trust, as has been observed in several cases~\cite{felez2022insider,Nelson@Coindesk,Solimano@AltLayer-TheDefiant}.
Ensuring transparency and fairness in airdrop design is therefore critical, not just for regulatory compliance, but for maintaining legitimacy in the eyes of the community.

\subsection{Summary of our Contributions}

We summarize our contributions as follows.

\StarCase{\stress{\textbf{Arbitrum study.}}}
We perform a comprehensive case study of the Arbitrum~\cite{Kalodner-Usenix} airdrop by measuring elements such as transaction volume, token distribution structure, and token value before and after the airdrop period.  We observed a notable increase in total daily fees coinciding with the airdrop event. However, the per-address transaction count for Arbitrum declined following the airdrop. Other protocols performed better than Arbitrum when no airdrop occurred. 

\StarCase{\stress{\textbf{Quantitative analysis.}}}
We conducted a quantitative analysis of airdrops executed by 1inch, Arbitrum, Arkham, DYDX, ENS, Lido, Optimism, Tornado Cash, and Uniswap.
We show that most of the funds received through these airdrops are sold on exchanges in the first airdropped token transfer made by the user after receiving the airdrop, instead of being used to create dapp or platform engagement on the part of end-users. This corresponds to 65.75\% of all Lido tokens, followed by 1inch with 58.67\%, and Optimism with 48.21\%, among others. On average, tokens were typically sold on exchanges within 1.17 to 2.76 transfers after being received, with a median of 1 to 2 transfers.

\StarCase{\stress{\textbf{Qualitative analysis.}}}
We conducted a qualitative analysis of past airdrops and propose guidelines for the development of future airdrops to mitigate the issues we identified. Our focus lies on airdrop farming, the distribution of governance tokens through airdrops, and insider trading. To address these concerns, we suggest alternative incentives, such as \stress{offering fee discounts for subsequent interactions within the blockchain protocol}.

\StarCase{\stress{\textbf{Multi-chain empirical data.}}}
We collected data from two major yet underexplored Ethereum rollups—Arbitrum and Optimism—as well as from nine major airdrops. We also gathered ZKsync Era data to discuss a fake airdrop performed by sybil farmers on ZKsync Era (see \S\ref{sec:gemstone}). Additionally, we curated address labels to identify centralized and decentralized exchange addresses. To promote reproducibility and support future research, we publicly release our dataset and data collection scripts in an open-access repository~\cite{Messias-DataSet-Code-2025}.

\begin{table*}[t]
\centering
\caption{List of popular airdrops analyzed in this paper. We include major airdrops across various blockchain ecosystems. Market caps are reported as of April 1\tsup{st}, 2025~\cite{coinmarketcap2025}. \blue{*}We analyze only the first round of the Optimism airdrop.}
\label{tab:airdrop-table}
\resizebox{\textwidth}{!}{%
\begin{tabular}{lcccccr}
\toprule
\thead{Airdrop} & \thead{First Claim Date} & \thead{Last Claim Date} & \thead{Blockchain} & \thead{Occurrence} & \thead{Project Type} & \thead{Market Cap (USD)} \\
\midrule
    1inch~\cite{bukov2020mooniswap} & Dec. 2020 & Still active &  Ethereum & Continuous & DEX & \$270.24M\\
    Arbitrum~\cite{Kalodner-Usenix} & Mar. 2023 & Sept. 2023 & Arbitrum One & Single & Rollup & \$1.56B\\
    Arkham~\cite{arkham} & Jul. 2023 & Still active & Ethereum & Continuous & Analytics & \$118.89M\\
    DYDX~\cite{juliano2018dydx} & Sept. 2021 & Still active & Ethereum & Continuous & DEX & \$68.71M
\\
    ENS~\cite{ENS} & Nov. 2021 & Aug. 2024 & Ethereum & Single & Name service & \$586.96M\\
    Lido~\cite{lido} & Jan. 2021 & Still active & Ethereum & Continuous & Staking & \$819.18M\\
    Optimism~\cite{optimism} & May 2022 & Jul. 2023 & OP Mainnet\blue{*} & Multi-round & Rollup & \$1.24B\\
    Tornado Cash~\cite{tornadocash} & Dec. 2020 & Aug. 2022 & Ethereum & Single & Mixer & \$39.68M\\
    Uniswap~\cite{adams2021uniswap} & Sept. 2020 & Still active & Ethereum & Continuous & DEX & \$3.9B\\
\bottomrule
\end{tabular}}
\end{table*}

\section{Airdrop Desiderata} \label{sec:airdrop-desidrata}

Airdrops are powerful tools for promoting protocols and acquiring users, attracting newcomers, and incentivizing existing users to engage with these protocols and their applications. They are widely utilized for these purposes, with numerous examples documented in the literature (see Table~\ref{tab:airdrop-table}). Protocols can create tokens and use airdrops to distribute them to users. For example, blockchain rollup solutions like Arbitrum, Optimism, and ZKsync Era, as well as DeFi applications such as Uniswap, 1inch, DYDX, and ENS have all employed airdrops.

Airdrops can manifest in various forms, with single-round and multi-round distributions being the most prevalent~\cite{allenCryptoAirdropsEvolutionary2023}. In a single-round airdrop, tokens are distributed to users all at once, whereas multi-round airdrops involve distributing tokens over multiple rounds each following different strategies. This approach allows leveraging insights from previous rounds to address challenges encountered, such as mitigating potential Sybil (i.e., multiple accounts controlled by a single entity) by observing past user behavior patterns~\cite{allenCryptoAirdropsEvolutionary2023}. The choice between single-round and multi-round airdrops depends on the protocol's objectives and community dynamics. 

Moreover, the timing of an airdrop can impact the number of eligible users receiving tokens. Established protocols opting for delayed airdrops may have a larger user base compared to newer projects executing airdrops sooner. This difference in user base size introduces complexity when performing Sybil detection, as there are more accounts to evaluate and potentially safeguard against. If not performed correctly, this could impact community happiness. In such instances, the protocol may inadvertently reward accounts linked to industrial farming, which could be viewed negatively by the community. 
To mitigate this issue, LayerZero Labs implemented a self-reporting mechanism for sybils~\cite{LayerZero-sybil-reporting}. Under this system, sybils have the option to self-report and receive 15\% of their designated token allocation.

% Airdrops are typically used to bootstrap a community of users.
Next, we decompose airdrop's high-level goal to bootstrap a community of users into multiple sub-goals.
These sub-goals are not necessarily orthogonal and there may be others that are worthwhile to pursue; we focus on them as they serve to illustrate interesting issues faced by common airdrop mechanisms.

\paraib{Attract users in the short-term}
Historically, airdrops have been employed by new blockchain protocols to establish an initial user base and to provide an initial boost of liquidity to the underlying chain and its protocols. Decentralized platforms, in particular, tend to become more useful and attractive to users as the level of economic activity increases.

\paraib{Foster long-term involvement}
Bootstrapping an initial base of users is important, but not enough for sustaining a high degree of economic activity over time.
Ideally, users should become \stress{regular}, everyday users.
This can be achieved by issuing rewards that can only be used within the blockchain protocol or application, similarly to frequent-flyer points.
For example, in \glspl{L2}~\cite{sguanci2021layer}, one may award fees discounts on future transactions.  
Other measures that may help are having multiple airdrops and rewarding users who perform ``quests'' that expose them to what the protocol has to offer.
For example, Linea's airdrop quests gave users an in-depth tour of their features and use-cases \cite{labs2023linea}.

\paraib{Target value-creating users}
An airdrop should ideally focus on users who can contribute the most to the long-term sustainability of their platform.
In the context of platforms that rely on user-based liquidity provisioning, this may refer to users who provide the most liquidity to lending pools~\cite{yaish2022blockchain} and \glspl{DEX}~\cite{yaish2023suboptimality}, or across a variety of tokens.
In rollups, one may consider so-called ``creators'' who deploy popular and useful contracts or users who bridge tokens to the rollup~\cite{ZKsync-value-scaling} as particularly valuable.
Such users provide additional use cases for a platform, and thus serve to attract other users.

\section{Data Collection}\label{sec:dataset}

\begin{table}[t]
\centering
\caption{Blockchain datasets collected for airdrop analysis.
Overview of the smart contract addresses, blockchain networks, event types, and time ranges used to collect token transfer and airdrop claim data for each protocol. For each dataset, we report the number of token transfers and airdrop claims identified between the specified start and end blocks.}
\label{tab:transfer_claim_transactions}
\resizebox{\textwidth}{!}{%
\begin{tabular}{llcccccccrr}
\toprule
\multirow{2}{*}{\thead{Protocol}} & \multirow{2}{*}{\thead{Address}} & \multirow{2}{*}{\thead{Chain}} & \multirow{2}{*}{\thead{Event type}} & \multirow{2}{*}{\thead{Start block}} & \multirow{2}{*}{\thead{End block}} & \multirow{2}{*}{\thead{Start date}} & \multirow{2}{*}{\thead{End date}} & & \multicolumn{2}{c}{\textbf{Number of}} \\
&&&&&&&&& \textbf{Transfers} & \textbf{Claims} \\
\midrule
1inch        & \href{https://etherscan.io/address/0x111111111117dc0aa78b770fa6a738034120c302}{0x1111$\cdots$c302} & Ethereum  & Transfer        & \num{11511393} & \num{21000000} & 2020-12-23 & 2024-10-19 & & \num{1782998} & \num{43806}  \\
Arbitrum     & \href{https://etherscan.io/address/0x912ce59144191c1204e64559fe8253a0e49e6548}{0x912c$\cdots$6548} \& \href{https://etherscan.io/address/0x67a24ce4321ab3af51c2d0a4801c3e111d88c9d9}{0x67a2$\cdots$c9d9} & Arbitrum & Transfer \& Claim        & \num{70398215} & \num{211336634} & 2023-03-16 & 2024-05-14 & & \num{47330340} & \num{583137}  \\
Arkham       & \href{https://etherscan.io/address/0x6e2a43be0b1d33b726f0ca3b8de60b3482b8b050}{0x6e2a$\cdots$b050} \& \href{https://etherscan.io/address/0x08c7676680f187a31241e83e6d44c03a98adab05}{0x08c7$\cdots$ab05} & Ethereum & Transfer \& Claim        & \num{17628655} & \num{20999848} & 2023-07-05 & 2024-10-19 & & \num{283775} & \num{64724}  \\
DYDX         & \href{https://etherscan.io/address/0x92d6c1e31e14520e676a687f0a93788b716beff5}{0x92d6$\cdots$eff5} \& \href{https://etherscan.io/address/0x01d3348601968ab85b4bb028979006eac235a588}{0x01d3$\cdots$a588} & Ethereum & Transfer \& RewardsClaimed        & \num{12809555} & \num{21000000} & 2021-07-12 & 2024-10-19 & & \num{747372} & \num{72870} \\
ENS          & \href{https://etherscan.io/address/0xc18360217d8f7ab5e7c516566761ea12ce7f9d72}{0xc183$\cdots$9d72} & Ethereum & Transfer \& Claim        & \num{13533418} & \num{20999950} & 2021-11-01 & 2024-10-19 & & \num{998619} & \num{102824}  \\
Lido         & \href{https://etherscan.io/address/0x5a98fcbea516cf06857215779fd812ca3bef1b32}{0x5a98$\cdots$1b32} & Ethereum & Transfer \& ClaimedTokens        & \num{11473276} & \num{20999995} & 2020-12-17 & 2024-10-19 & & \num{1107586} & \num{516}  \\
Optimism     & \href{https://etherscan.io/address/0x4200000000000000000000000000000000000042}{0x4200$\cdots$0042} \& \href{https://etherscan.io/address/0xfedfaf1a10335448b7fa0268f56d2b44dbd357de}{0xfedf$\cdots$57de} & Optimism & Transfer \& Claim           & \num{6491116}  & \num{117863006} & 2022-04-26 & 2024-03-24 & & \num{43990557} & \num{160603}  \\
Tornado Cash & \href{https://etherscan.io/address/0x77777feddddffc19ff86db637967013e6c6a116c}{0x7777$\cdots$116c}  & Ethereum  & Transfer        & \num{11474599} & \num{20999592} & 2020-12-18 & 2024-10-19 & & \num{278455} & \num{5474}  \\
Uniswap      & \href{https://etherscan.io/address/0x1f9840a85d5af5bf1d1762f925bdaddc4201f984}{0x1f98$\cdots$f984} & Ethereum  & Transfer           & \num{10861674} & \num{20999993} & 2020-09-14 & 2024-10-19 & & \num{4872925} & \num{221087}  \\
\bottomrule
\end{tabular}%
}
\end{table}

To analyze airdrop distribution patterns and their post-distribution market behavior (\S\ref{sec:analysis}), we compiled a comprehensive dataset from multiple blockchains and external data sources. Specifically, we collected data from Ethereum, Arbitrum, and Optimism using blockchain nodes, along with metadata from public sources such as Growthepie and address-labeling services.

\paraib{On-chain data collection}
Our primary dataset consists of smart contract event logs data, since their inception to October 2024, related to airdrop claims and token transfers for nine major protocols: 1inch, Arbitrum, Arkham, DYDX, ENS, Lido, Optimism, Tornado Cash, and Uniswap. For each project, we identified and extracted all relevant contract events, including airdrop eligibility, claim transactions, and subsequent token transfers. Table~\ref{tab:transfer_claim_transactions} lists the smart contracts and event types analyzed in this study. Some protocols—such as Arbitrum, Arkham, DYDX, and Optimism—use separate contracts for their usual token transfers and airdrop distribution. Others, including 1inch, ENS, Lido, Tornado Cash, and Uniswap, rely on a single contract to handle both token transfers and airdrop claims, typically distinguished by different event types (e.g., a \stress{Claim} and \stress{Transfer} events). In certain cases, such as 1inch, Tornado Cash, and Uniswap, the contract does not emit distinct events for \stress{Transfer} and \stress{Claim}. In these situations, we identify airdropped tokens by detecting transfers originating from known airdrop distribution addresses belonging to the respective projects.

\paraib{Exchange and address labeling}
To differentiate between regular users and exchange-associated addresses (both centralized and decentralized exchanges), we aggregated public address labels from Etherscan~\cite{Etherscan@ETH-explorer}, Optimism Explorer~\cite{Optimism@explorer}, Arbiscan~\cite{Arbitrum@explorer}, Dune~\cite{hildobby@Dune}, and similar services. Additionally, we queried the Arkham Intelligence API~\cite{arkham} for high in-degree nodes—i.e., addresses that frequently receive tokens from many accounts—which are characteristic of exchange wallets. With these approaches, we labeled the top 150 such addresses for each of the nine protocols and ultimately identified a total of 3569 exchange addresses: 3011 centralized (CEX) and 558 decentralized (DEX).

\paraib{Token distribution visualization}
For the token distribution graphs presented in \S\ref{subsec:token_distribution_graphs}, we used the open-source tool Gephi to visualize address interactions and airdrop flows. To enhance the readability of the network visualizations, we focused on a subgraph consisting of each account's first token transfer following the airdrop receipt.

\paraib{Cross-chain ecosystem data}
In \S\ref{sec:status-quo}, we complement our analysis with macro-level blockchain metrics from Growthepie.xyz~\cite{Growthepie} for the year 2023. These include daily transaction counts, active addresses, median transaction fees, \gls{TVL}, user fees, and stablecoin market capitalizations across leading Layer-2 ecosystems (ZKsync Era, OP Mainnet, Arbitrum, Immutable X, and Polygon zkEVM). All time series were smoothed using a 7-day moving average to enhance interpretability.

\paraib{Reproducibility}
\stress{To ensure the reproducibility of our results, we publicly release the complete dataset and analysis scripts in an open-access repository~\cite{Messias-DataSet-Code-2025}.}

\section{Exchange Aftermarket Analysis}
\label{sec:analysis}

The motivation for our quantitative analysis in this paper comes from the following observation: \stress{airdrop recipients quickly sell tokens and jump ship}, which obviously defeats the point of airdrops in the first place.
Analyses performed on \gls{DEX} airdrops show that airdrop recipients sometimes sell all of their tokens shortly after receiving them.
For example, shortly after ParaSwap's airdrop, 61\% of their tokens were sold~\cite{fanAltruisticProfitorientedMaking2023}.

In both cases, the majority of recipients ceased using the relevant blockchain protocols within a few months. This pattern suggests that the airdrop was ineffective in keeping recipients engaged, or that many recipients were likely Sybil accounts. Moreover, a rapid sell-off could unsettle the market, particularly if seen as an indication of dwindling confidence in the protocol's future prospects. In our study, we examine data associated with nine airdrops presented in Table~\ref{tab:airdrop-table}, which were gathered from blockchain data of Ethereum, Arbitrum, Optimism, and ZKsync Era (as discussed in \S\ref{sec:dataset}).

Interestingly, some addresses associated with DYDX, Arkham, Lido, and Tornado Cash received airdropped tokens multiple times. This could be attributed to fees, yield payments, or other protocol distributions. However, per Table~\ref{tab:airdrop-table}, this is in line with their airdrop design that is \stress{continuous}, meaning that the same address could qualify for the next airdrop round.
In fact, DYDX addresses received an airdrop on average 1.58 times, with a standard deviation of 1.68, ranging from 1 to 30 instances, and a median of 1.
For Arkham, five addresses received the airdrop twice. Similarly, nine addresses linked to Lido also received two airdrops.
In the case of Tornado Cash, 99 addresses received airdrops more than once, with an average of 1.02 instances, a standard deviation of 0.26, a median of 1, and a range from 1 to 16 occurrences.

\subsection{Tokens Distributed and Liquidated Over Time}
\label{sec:token_distribution_liquidation}

We analyze how airdropped tokens were distributed and what actions recipients took following their receipt. We organize our findings into four key areas: claim timing, early liquidation behavior, interaction with exchanges, and continued protocol engagement.

\subsubsection{Airdrop Claim Behavior}
\begin{figure}[t]
\centering
\centering
\includegraphics[width=\onecolgrid]{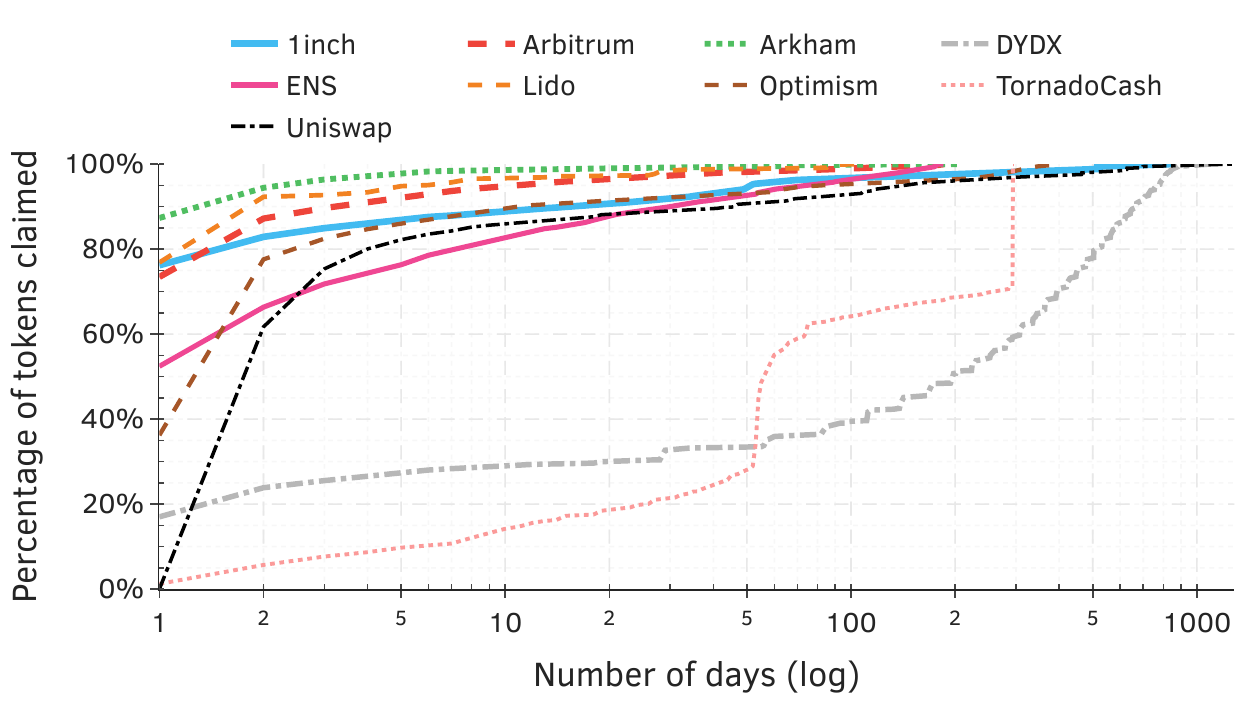}
\caption{Daily token claims made by accounts eligible to receive an airdrop.}
\label{fig:daily-token-claim}
\end{figure}

Figure~\ref{fig:daily-token-claim} illustrates the rate at which eligible users claimed their airdropped tokens. The Arkham, Lido, and Arbitrum airdrops saw the fastest and most concentrated activity: 87\%, 77\%, 73\%, respectively, of eligible accounts claimed tokens on the first day, and nearly 94\%, 92\%, and 87\%, respectively, by the second day. This pattern suggests strong anticipation or automation, likely driven by airdrop farming tools or pre-announced eligibility windows. Other protocols, such as ENS, DYDX, and Uniswap, exhibited more gradual claiming behavior, with long tails spanning multiple weeks.

\subsubsection{Initial Liquidation: Airdrops Sold Shortly After Receipt}
Table~\ref{tab:distribution-stats} shows the percentage of tokens sold at exchanges as recipients' first action post-claim (column 4). Lido had the highest liquidation rate (65.75\%), followed by 1inch (58.67\%) and Optimism (48.21\%). In contrast, DYDX and Arkham experienced lower immediate liquidation rates at 13.56\% and 19.79\%, respectively. This variation likely stems from protocol-specific factors such as token utility, incentive structures, or community expectations.

\begin{figure}[t]
\centering
\begin{subfigure}{\twocolgrid}
    \centering
    \includegraphics[width=\twocolgrid]{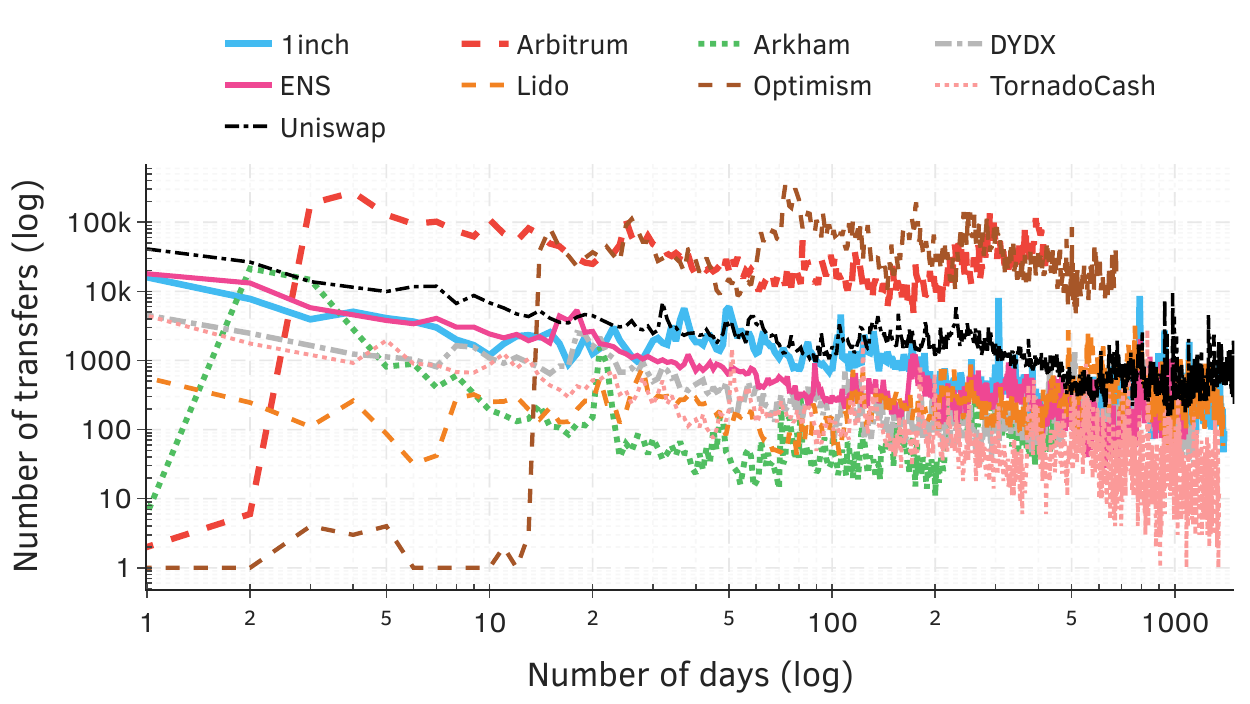}
    \caption{All accounts.}
    \label{fig:daily-token-transfers-to-exchanges-all-accounts}
\end{subfigure}
\begin{subfigure}{\twocolgrid}
    \centering
    \includegraphics[width=\twocolgrid]{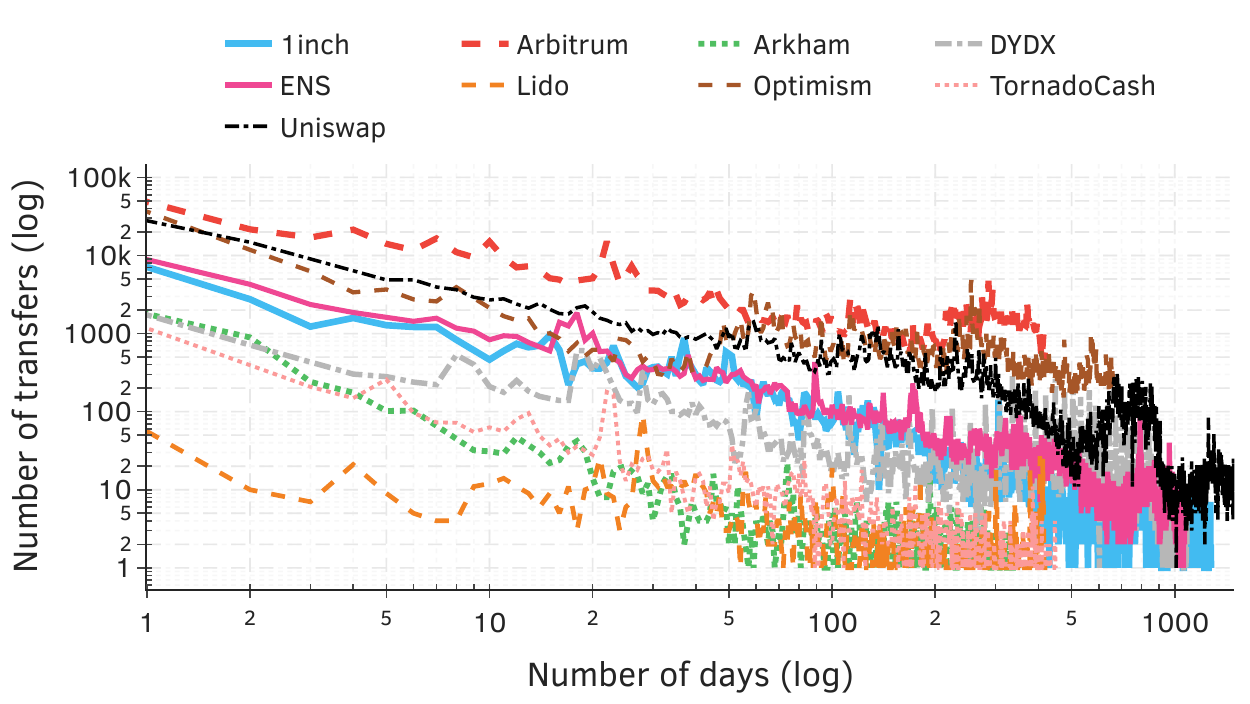}
    \caption{Accounts with airdropped tokens.}
    \label{fig:daily-token-transfers-to-exchanges-airdropped-tokens}
\end{subfigure}
\caption{Comparison of daily token transfer to exchanges: (a) all accounts; and (b) only those that received an airdrop.}
\label{fig:token-transfer-to-exchange}
\end{figure}

\begin{figure}[t]
\centering
\begin{subfigure}{\twocolgrid}
    \centering
    \includegraphics[width=\twocolgrid]{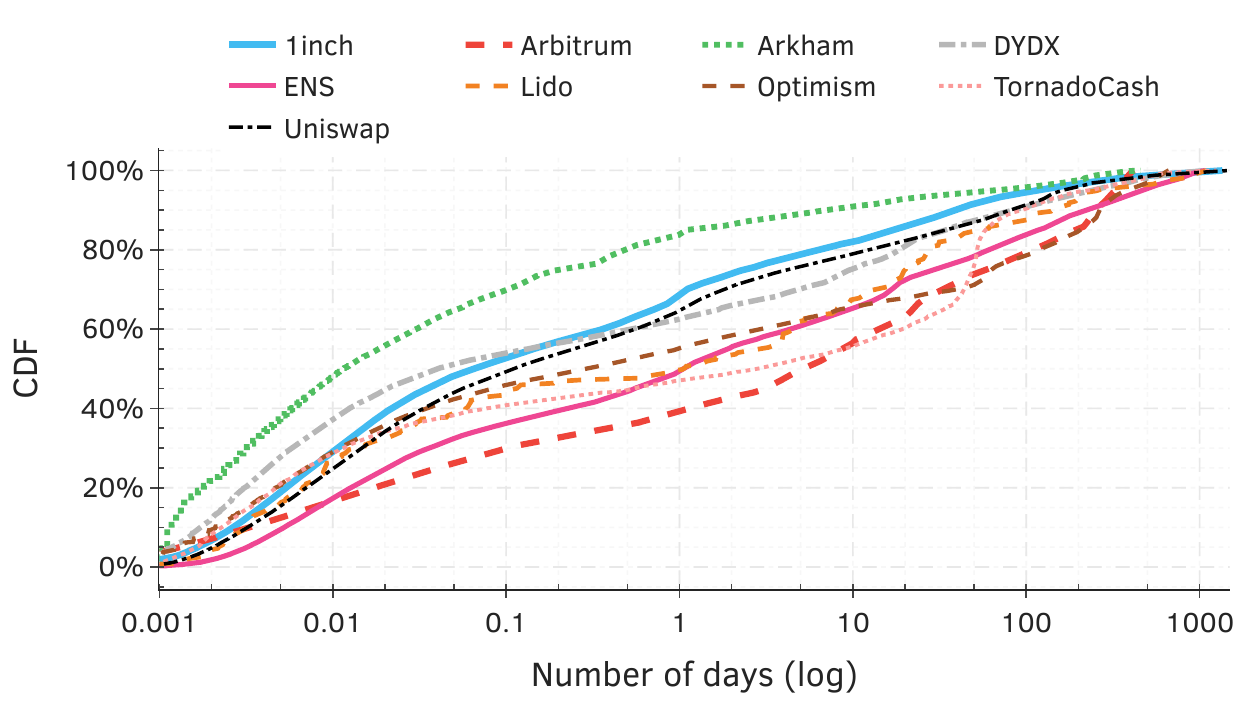}
    \caption{Time to reach an exchange.}
    \label{fig:cumulative-over-days}
\end{subfigure}
\begin{subfigure}{\twocolgrid}
    \centering
    \includegraphics[width=\twocolgrid]{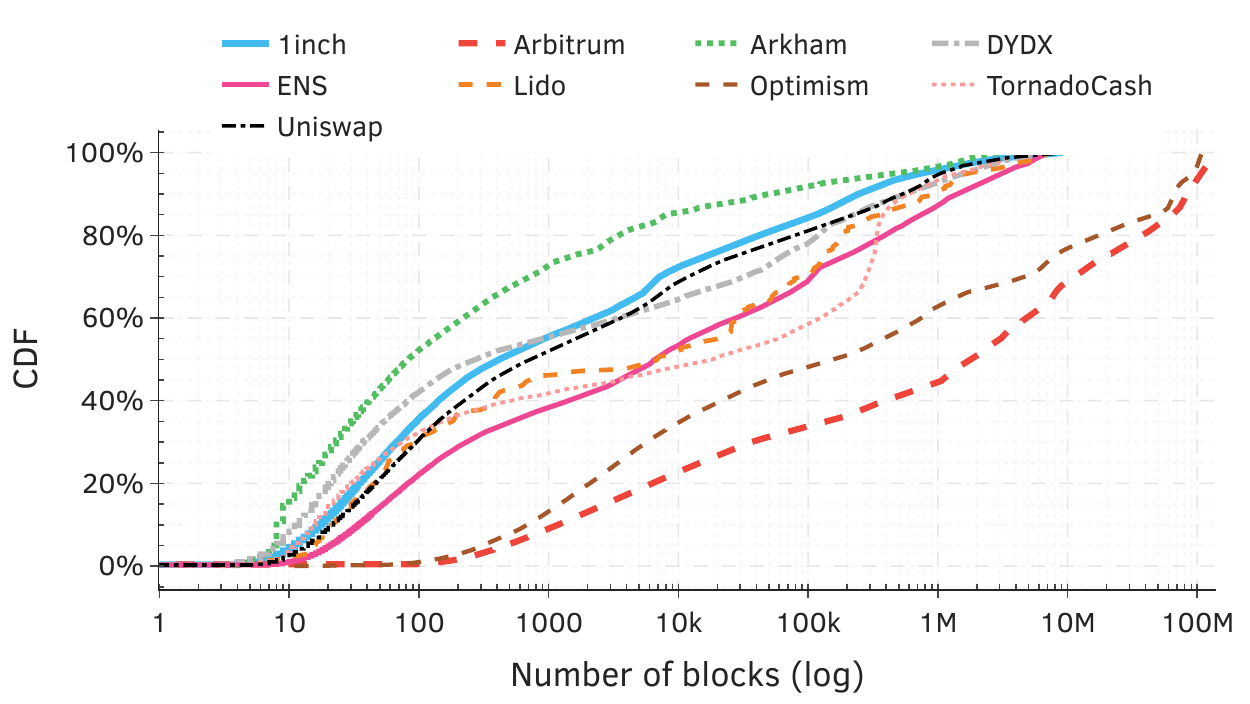}
    \caption{Number of blocks to reach an exchange.}
    \label{fig:cumulative-over-blocks}
\end{subfigure}
\caption{Comparison of token claiming and transfer patterns post-airdrop: (a) time taken to reach an exchange; and (b) number of blocks to reach an exchange.}
\label{fig:cumulative-until-exchange}
\end{figure}

These findings are supported by Figures~\ref{fig:token-transfer-to-exchange} and~\ref{fig:cumulative-until-exchange}, which reveal the timing and structure of token transfers. Specifically, Figure~\ref{fig:cumulative-over-days} shows that 66.09\% of 1inch recipients transferred tokens to exchanges within 24 hours. Similarly, ENS and Lido saw over half of their recipients interact with exchanges in the same timeframe. Figure~\ref{fig:cumulative-over-blocks} indicates the results based on the number of blocks. The frequency at which blocks are committed varies according to the evaluated blockchain. That is, Ethereum processes blocks at a much lower pace than Arbitrum, Optimism, and ZKsync Era that are \gls{L2} solutions.

\begin{table}[t]
\centering
\caption{Airdrop distribution statistics for the nine airdrops as of October 2024. We also report the percentage of tokens sold at exchanges by the accounts as their first ever transfer after receiving the airdrop. Note that protocols often send a high share of airdropped tokens to addresses they control. For more details on the top earners, refer to Table~\ref{tab:top-airdrop-earners} in \S\ref{sec:top_airdrop_earners}.}
\resizebox{\textwidth}{!}{%
\begin{tabular}{lcccrrrrr}
\toprule
 \multirow{2}{*}{\thead{Protocol}}  & \multicolumn{1}{c}{\thead{Token}}   & \multicolumn{1}{c}{\thead{Tokens}}   & \multicolumn{1}{c}{\thead{\% sold at}}   & \multicolumn{1}{c}{\thead{\# of recipient}}   &  \multicolumn{4}{c}{\thead{Tokens Received per Address}}     \\
 &  \thead{Supply} & \thead{Distributed}  & \thead{exchanges} & \thead{addresses} &   \thead{Mean} & \thead{Median} & \thead{Min} & \thead{Max} \\
\midrule
    1inch & $1.5\times10^{9}$ & $83.6\times10^{6}$   & 58.67\% &  \num{43806} &  \num{1909.39} & \num{636.44} & \num{0} & $9.75\times10^{6}$ \\
    Arbitrum & $10\times10^{9}$ & $1.09\times10^{9}$  & 28.64\% &  \num{583137} &  \num{1874.02} & \num{1250} & \num{625} & \num{10250} \\ 
    Arkham & $1\times10^{9}$ & $29.39\times10^{6}$  & 19.79\% &  \num{64724} &  \num{454.15} & \num{197.43} & \num{1.97} & \num{249353.40} \\ 
    DYDX & $1\times10^{9}$ & $167.51\times10^{6}$  & 13.56\% &  \num{72870} &  \num{2298.77}  & \num{562.24} & \num{0} & $1.49\times10^{6}$ \\
    ENS &  $100\times10^{6}$ & $19.6\times10^{6}$  & 46.74\% &  \num{102824} &  \num{190.93} & \num{152.4} & \num{0.0001} & \num{1143.54} \\
    Lido & $1\times10^{9}$ & $3.97\times10^{6}$  & 65.75\% &  \num{516} &  \num{7695.82} & \num{238.78} & 0.000871 & \num{746286.49} \\ 
    Optimism & $\approx 4.29\times10^{9}$ & $166.25\times10^{6}$  & 48.21\% &  \num{160603} &  \num{1035.17} & \num{776.87} & \num{0} & \num{32431.80} \\ 
    Tornado Cash & $\approx 1\times10^{9}$ & $500\times10^{3}$  & 33.71\% &  \num{5474} &  \num{91.34} & \num{21.77} & \num{0} & \num{143831.16} \\ 
    Uniswap & $1\times10^{9}$ & $137\times10^{6}$  & 43.69\% &  \num{221087} &  \num{619.85} & \num{400} & \num{400} & $2.1\times10^{6}$ \\
     \bottomrule
\end{tabular}
}
\label{tab:distribution-stats}
\end{table}

\subsubsection{Exchange Interaction and Transfer Hops}
To analyze the path tokens take from airdrop recipients to CEXs and DEXs, we consider a blockchain transaction graph and formally define the number of ``hops'' between a source node (i.e., an airdrop recipient) and a terminal node (i.e., an exchange).

\begin{definition}[Number of hops to reach an exchange]
Let \( \mathcal{G} = \left(\mathcal{V}, \mathcal{E} \right) \) denote a graph where $\mathcal{V}$ is the set of all blockchain addresses and an edge $(u,v) \in \mathcal{E}$ represents a transfer from address $u$ to address $v$.
Furthermore, let $\mathcal{R} \subset \mathcal{V}$ be the set of airdrop recipient addresses and $\mathcal{D} \subset \mathcal{V}$ be the set of exchange addresses.
For any two addresses $u,v \in \mathcal{V}$, denote by $\text{dist}(u,v)$ the length of the shortest path from $u$ to $v$ in $\mathcal{G}$.
If no path exists, then $\text{dist}(u,v) = \infty$.

Given an airdrop recipient $r \in \mathcal{R}$, the \textit{number of hops to reach an exchange} \( H(r) \) is defined as the shortest path from $r$ to any exchange in $\mathcal{X}$:
\[
H(r) = \min_{x \in \mathcal{X}} \left( \text{dist}(r,x) \right)
\]
\end{definition}

Table~\ref{tab:interaction_with_exchange} quantifies how directly recipients sent tokens to exchanges. We find that most addresses liquidated tokens within 1--2 hops: for example, the median number of hops for Arbitrum, Arkham, and DYDX was 2, while others had medians of 1. The fraction of recipients who interacted with exchanges was substantial: over 60\% for 1inch and Optimism, and over 75\% for Lido. Only a few protocols, like Arkham (15.63\%), had lower exchange interaction rates, likely due to token holding incentives or smaller allocations.

These interactions were not limited to simple sell-offs. As shown in Table~\ref{tab:distribution_days}, the median number of days to transfer to an exchange was almost under 1 for nearly all protocols. For instance, Arkham and DYDX saw medians of 0.01 and 0.04 days, respectively. Despite some outliers with longer delay times, the heavy left tail in Figure~\ref{fig:cumulative-until-exchange} indicates that a large majority of sales occurred almost immediately after claim.

\begin{table}[t]
\centering
\caption{Interaction between airdrop recipients and exchanges, measured by the total number of exchange interactions per protocol and the number of hops needed to reach an exchange. Higher median hop counts, seen in DYDX, Arbitrum, and Arkham, suggest more complex paths before tokens reach exchanges. In contrast, 1inch, Lido and Tornado Cash exhibit shorter paths while Arbitrum, Optimism, and Uniswap higher direct interaction rates, suggesting more frequent exchange activities by their airdrop recipients.}
\resizebox{\textwidth}{!}{%
\begin{tabular}{lrrcrccrcccccc}
\toprule
\multirow{2}{*}{\thead{Protocol}}  & \multicolumn{3}{c}{\thead{Airdrop recipients}} & \multicolumn{4}{c}{\thead{\# of interactions with exchange}} & \multicolumn{4}{c}{\thead{\# of hops to reach an exchange}} \\
& \thead{Total} & \thead{\# sent to exchanges}  & \thead{\% sent to exchanges} & \thead{Mean} & \thead{Median} & \thead{Min.} & \thead{Max.} & \thead{Mean} & \thead{Median} & \thead{Min.} & \thead{Max.}\\ 
\midrule
{1inch} & \num{43806}   & \num{26389} & 60.24\% & 2.64 & 1 & 1 & 2695& 1.42   & 1  & 1  & 7  \\
{Arbitrum} & \num{583137}   & \num{235346} & 40.36\% & 4.11  & 1 & 2 & \num{19336} & 1.69 & 2  & 1  & 31  \\
{Arkham} & \num{64724}   & \num{10116} & 15.63\% & 1.05 & 1 & 1 & 14 & 2.76 &  2  & 1  & 79  \\
{DYDX} & \num{72870}   & \num{16383} & 22.48\% & 2.39 & 1 & 1 & 196 & 1.76 &  2  & 1  & 17  \\
{ENS} & \num{102824}   & \num{53717} & 52.24\% & 1.78 & 1 & 1 & 107 & 1.49 &  1  & 1 & 50  \\
{Lido} & \num{516}   & \num{395} & 76.55\% & 4.21 & 1 & 2 & 132 & 1.17 &  1  & 1  & 4  \\
{Optimism} & \num{160603}   & \num{98168} & 61.12\% & 4.71 & 1 & 1 & \num{16065} & 1.47 &  1  & 1  & 32  \\
{TornadoCash} & \num{5474}   & \num{2951} & 53.91\% & 2 & 1 & 1 & 53 & 1.55 &  1  & 1  & 5  \\
{Uniswap} & \num{221087}   & \num{118880} & 53.77\% & 3.10 & 1 & 1 & \num{63409} & 1.56 & 1  & 1  & 16  \\
\bottomrule
\end{tabular}
}
\label{tab:interaction_with_exchange}
\end{table}

\begin{table}[t]
\centering
\caption{Distribution analysis of how long took accounts to send tokens to exchanges. For each protocol, we report the number of airdrop recipients, how many transferred tokens to an exchange, and the share of such users. We also provide descriptive statistics (in days) on the time taken to move tokens to an exchange, including mean, standard deviation, minimum, median, and maximum values.}
\resizebox{\textwidth}{!}{%
\begin{tabular}{lrrcrrccr}
\toprule
\multirow{2}{*}{\thead{Protocol}}  & \multicolumn{3}{c}{\thead{Airdrop recipients}} & \multicolumn{5}{c}{\thead{\# of days to transfer to an exchange}}  \\
& \thead{Total} & \thead{\# sent to exchanges}  & \thead{\% sent to exchanges} & \thead{Mean} & \thead{Std.} & \thead{Min.} & \thead{Median.} & \thead{Max.} \\ 
\midrule
{1inch} & \num{43806}   & \num{26389} & 60.24\% & 24.96 & 107.15 & 0 & 0.07 & 1352.73    \\
{Arbitrum} & \num{583137}   & \num{235346} & 40.36\% & 61.46 & 107.00 & 0 & 5.13 & 418.37  \\
{Arkham} & \num{64724}   & \num{10116} & 15.63\% & 11.37 & 47.71 & 0 & 0.01 & 456.76   \\
{DYDX} & \num{72870}   & \num{16383} & 22.48\% & 37.09 & 119.32 & 0 & 0.04 & 1091.45  \\
{ENS} & \num{102824}   & \num{53717} & 52.24\% & 69.36 & 172.97 & 0 & 1.04 & 1073.65 \\ 
{Lido} & \num{516}   & \num{395} & 76.55\% & 55.32 & 167.47 & 0 & 1.08 & 1099.55   \\
{Optimism} & \num{160603}   & \num{98168} & 61.12\% & 70.24 & 137.44 & 0 & 0.30 & 662.77  \\
{TornadoCash} & \num{5474}   & \num{2951} & 53.91\% & 46.06 & 114.31 & 0 & 2.86 & 1248.52   \\
{Uniswap} & \num{221087}   & \num{118880} & 53.77\% & 32.27 & 119.33 & 0 & 0.11 & 1490.03   \\
\bottomrule
\end{tabular}
}
\label{tab:distribution_days}
\end{table}

\subsubsection{Recipient Engagement and Token Utilization}
To assess long-term engagement, we studied token transfer patterns post-airdrop using Table~\ref{tab:distribution_transfers}. We found that while over 60\% of airdrop recipients across all protocols performed at least one transfer, most did not continue interacting beyond the initial transaction.

For example, the median number of token-related transfers per recipient was one across most protocols, with only Lido (median = 2) exceeding this marginally. Likewise, the time span between a recipient's first and last transfer (a proxy for continued engagement) was short: the median was 0 days for most protocols. Only Lido exhibited a higher median of 42.62 days, reflecting longer engagement likely due to staking or yield incentives.

\begin{figure}[t]
\centering
\begin{subfigure}{\twocolgrid}
    \centering
    \includegraphics[width=\twocolgrid]{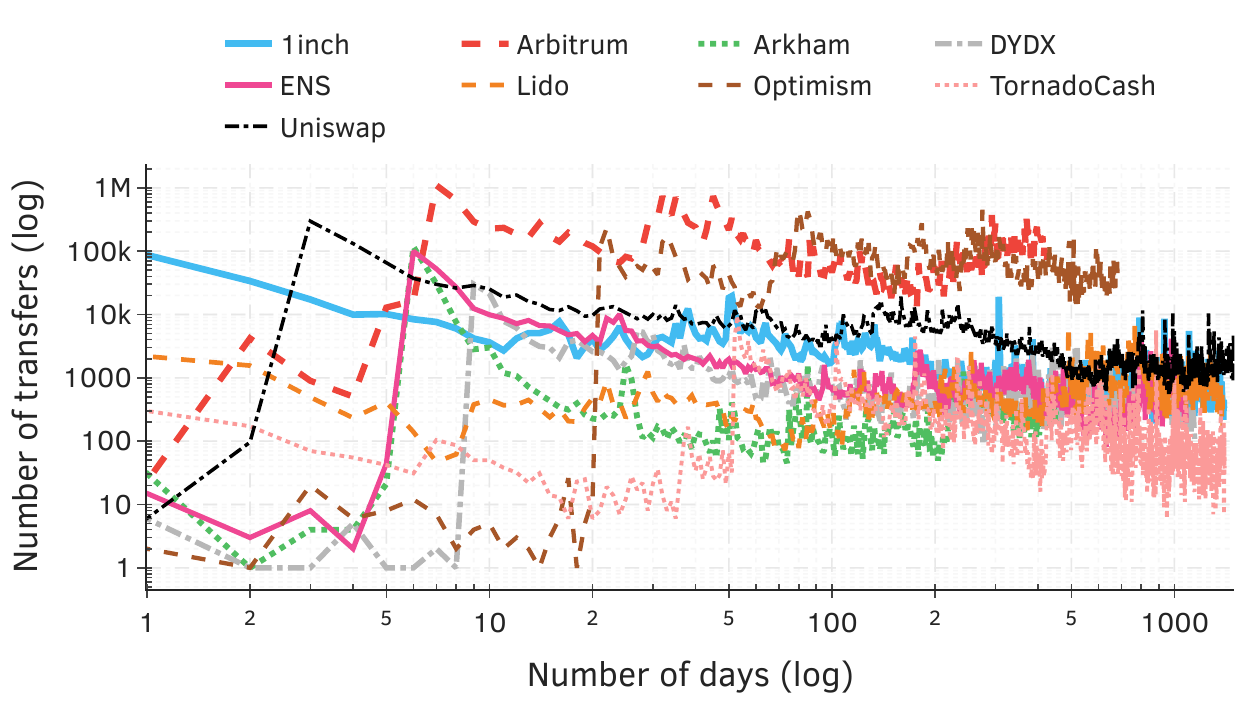}
    \caption{All accounts.}
    \label{fig:daily-token-transfers-all}
\end{subfigure}
\begin{subfigure}{\twocolgrid}
    \centering
    \includegraphics[width=\twocolgrid]{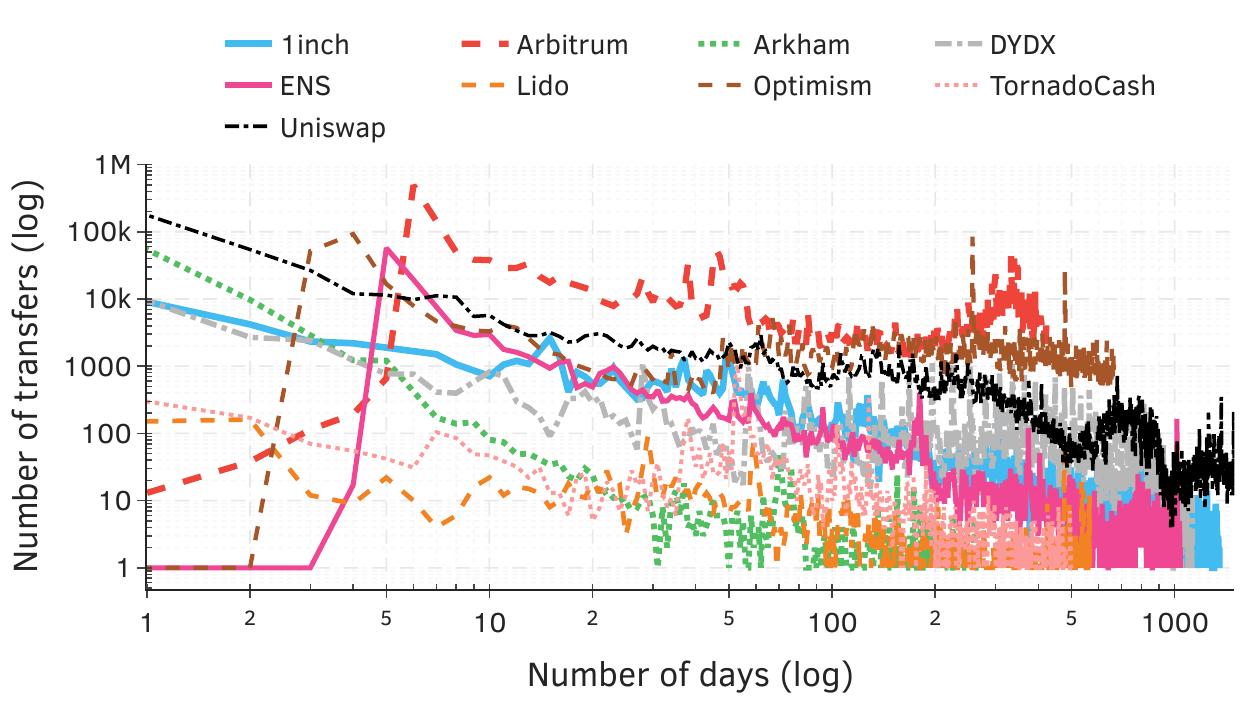}
    \caption{Accounts with airdropped tokens.}
    \label{fig:daily-token-transfers-airdropped-accounts}
\end{subfigure}
\caption{Comparison of token number of token transfers made: (a) by all accounts; and (b) only by those that received an airdrop.}
\label{fig:daily-token-transfer}
\end{figure}

These findings are consistent with the activity shown in Figure~\ref{fig:token-transfer-to-exchange} and Figure~\ref{fig:daily-token-transfer}. Specifically, recipients' token transfer activity peaked within days of the airdrop and dropped sharply thereafter, indicating that tokens were often liquidated in a one-time event rather than incorporated into ongoing protocol usage.

\begin{table}[t]
\centering
\caption{Post-airdrop token activity and account lifespan across protocols. It shows the share of airdrop recipients who engaged in at least one token transfer, along with summary statistics on their transfer behavior. For each protocol, we report the average and median number of transfers per address, as well as the duration between the first and last token-related activity. A subset of recipients never initiated a transfer, potentially indicating lost access, strategic holding, or disinterest in participation.}
\label{tab:distribution_transfers}
\resizebox{\textwidth}{!}{%
\begin{tabular}{lrcrccrrccrr}
\toprule
\multirow{2}{*}{\thead{Protocol}}  & \multirow{2}{*}{\thead{Total of accounts}} & \multicolumn{5}{c}{\thead{\# of transfers per address}} & \multicolumn{5}{c}{\thead{Days between first and last token transfers}} \\
&  & \thead{Mean} & \thead{Std.}  & \thead{Min.} & \thead{Median} & \thead{Max.} & \thead{Mean} & \thead{Std.}  & \thead{Min.} & \thead{Median} & \thead{Max.}\\ 
\midrule
1inch & \num{42759} (97.61\%) & 2.55 & 17.54 & 1 & 1 & \num{3195} & 62.76 & 190.18 & 0 & 0 & \num{1388.78}\\
Arbitrum & \num{570450} (97.82\%) & 4.30 & 89.45 & 1 & 1 & \num{40771} & 86.65 & 136 & 0 & 0 & 418.43\\
Arkham & \num{64032} (98.94\%) & 1.06 & 0.42 & 1 & 1 & 58 & 2.93 & 27.84 & 0 & 0 & 458.57\\
DYDX & \num{44773} (97.45\%) & 2.20 & 3.52 & 1 & 1 & 214 & 84.38 & 198.32 & 0 & 0 & \num{1136.17}\\
ENS & \num{93506} (90.94\%) & 1.70 & 1.83 & 1 & 1 & 117 & 57.79 & 175.26 & 0 & 0 & \num{1074.29} \\
Lido & \num{471} (92.9\%) & 4.33 & 8.90 & 1 & 2 & 147 & 231.17 & 348.81 & 0 & 42.62 & \num{1348.97}\\
Optimism & \num{154209} (96.02\%) & 4.83 & 51.79 & 1 & 1 & \num{16089} & 163.63 & 222.20 & 0 & 0 & 663.21\\
Tornado & \num{4900} (91.64\%) & 1.85 & 5.59 & 1 & 1 & 327 & 20.17 & 92.57 & 0 & 0 & \num{1321.94}\\
Uniswap & \num{217635} (98.44\%) & 3.73 & 228.29 & 1 & 1 & \num{84328} & 64.93 & 195.98 & 0 & 0 & \num{1493.33}\\

\bottomrule
\end{tabular}}
\end{table}

\subsubsection{Top Airdrop Earners} \label{sec:top_airdrop_earners}

In this section, we present the top-1 airdrop recipient for each of the nine protocols analyzed. Airdropped tokens may be distributed to various types of accounts, including \glspl{EOA}, smart contracts, and multi-signature (multisig) wallets. Some of these recipients might be directly affiliated with the protocol—such as treasury wallets or governance-related accounts—while others may represent independent users or airdrop farming operations.

Multisig wallets, which require multiple signatures to authorize transactions, are often used by protocol teams for treasury or operational purposes~\cite{messias2023understanding}. Consequently, top recipients frequently include protocol-controlled accounts, as discussed in Table~\ref{tab:distribution-stats}. 

Table~\ref{tab:top-airdrop-earners} lists the top airdrop recipients for each protocol, along with the number and percentage of tokens received, both in relation to the airdrop and the total token supply. Wallet labels were assigned using the methodology described in \S\ref{sec:dataset}. We highlight two representative examples. First, the wallet labeled \stress{analytico.eth} received \num{11.66}\% of the total tokens distributed by the 1inch airdrop. According to Etherscan, this address holds approximately \$7.5 million USD in tokens as of May 2025~\cite{Analytico@Etherscan} and has participated in at least 12 other airdrops~\cite{Analytico@Dune}. These patterns suggest that it is likely part of a coordinated airdrop farming operation. However, the identity behind this wallet remains unknown, and it is unclear whether it is operated by an individual or a group. In contrast, \num{28.77}\% of Tornado Cash airdropped tokens were distributed to the Tornado Cash Governance, a common practice that aligns with protocols allocating tokens to support long-term ecosystem development. Although some of these wallets received a significant share of the distributed tokens, their allocations remain relatively small when measured against the total token supply, underscoring the limited concentration of airdrops at the protocol level.

\begin{table}[t]
\centering
\caption{Top airdrop earner for the nine protocols analyzes in this paper: 1inch, Arbitrum, Arkham, DYDX, ENS, Lido, Optimism, Tornado Cash, and Uniswap.}
\resizebox{\textwidth}{!}{%
\begin{tabular}{llcrrr}
\toprule
\thead{Protocol} & \thead{Wallet address} & \thead{Label} & \thead{\# of tokens} & \thead{\% of dist. tokens} & \thead{\% of total supply} \\
\midrule
1inch & \href{https://etherscan.io/tx/0xb9cdd3c726984ce876ade2581a3cf130e5fd4eb9d14a85b157d43e5abf3c155e\#eventlog}{0xa0f7$\cdots$9b73} & analytico.eth & \num{9749686} & \num{11.66} & \num{0.65} \\
Arbitrum & \href{https://arbiscan.io/tx/0xe12e4da6ea6674078b9a78ebda4b422f94050aa453e32492dd70aa9defae2ace\#eventlog}{0x79f0$\cdots$2b3f} & ``luckboxcolonel'' on OpenSea & \num{10250} & \num{0.001} & \num{0.0001} \\
Arkham & \href{https://etherscan.io/tx/0x161cbff62dc298ef8083e0373b6619bc8c67ab7a1b1203504400b78c3f5a3629\#eventlog}{0xf0df$\cdots$8d2f} & 
Laura Kornelija
 & \num{249353.40} & \num{0.85} & \num{0.25} \\
DYDX & \href{https://etherscan.io/tx/0x7c9e8a62b1f1b5dea00cf9321b4020c1ff4ec0f235d72a8d9d081622194a6db4\#eventlog}{0xa615$\cdots$2d92} & --- & \num{1492217.38} & \num{0.89} & \num{0.15} \\
ENS & \href{https://etherscan.io/tx/0x57458e828809dc0efa892a507310cc1817151c2f6594edfa7f4a1fa6ab070181\#eventlog}{0x0904$\cdots$9859} & ENS wallet  & \num{1143.54} & \num{0.006} & \num{0.0011} \\
Lido & \href{https://etherscan.io/tx/0x615e59f5aa5a4c6f29156fe137c9679ca2f2e3bc05ca3f2e76c13f9ef99fb4db\#eventlog}{0x9796$\cdots$7632} & --- & \num{746286.49} & \num{18.79} & \num{0.75} \\
Optimism & \href{https://optimistic.etherscan.io/tx/0x39dc59ec07f2d52e3aacd05e9c2cdedb3e4771882ccfc4767d2f19438493a0f4\#eventlog}{0xba74$\cdots$ae07} & tarrence.eth & \num{32431.80} & \num{0.020} & \num{0.0008}\\
Tornado Cash & \href{https://etherscan.io/tx/0x57a2a742169870301727d760a1b91613a8407287a5c300800a2cd4bce560c258\#eventlog}{0x5efd$\cdots$a1ce} & Tornado Cash Governance & \num{143831.16} & \num{28.77} & \num{1.44} \\
Uniswap & \href{https://etherscan.io/tx/0xe109065efd962491ba9ae780a6d7e05fd0fcdfbdf066a68ee79027449f121c90\#eventlog}{0xdd27$\cdots$c3b3} & Sybil Delegate~\cite{Addresses@Sybil} & $2.1\times10^{6}$ & \num{1.54} & \num{0.21} \\
\bottomrule
\end{tabular}
}
\label{tab:top-airdrop-earners}
\end{table}

\subsection{Token Distribution Graphs}\label{subsec:token_distribution_graphs}

To better understand how airdropped tokens propagate across users, we construct and analyze token transfer networks for each protocol. We define a \textit{token distribution graph} as a multidirected graph \( G = (V, E) \), where each node \( v \in V \) represents a unique address, and each directed edge \( (u, v) \in E \) denotes a transfer of tokens from address \( u \) to address \( v \) after the airdrop.

These networks vary considerably in scale. The graph for 1inch consists of \num{370372} nodes and \num{1739191} edges, while Arbitrum's network includes \num{2789879} nodes and \num{46747201} edges. Arkham's network contains \num{278987} nodes and \num{467472} edges, and DYDX's has \num{181000} nodes and \num{658614} edges. The ENS network comprises \num{229991} nodes and \num{895792} edges, Lido's has \num{127754} nodes and \num{1107066} edges, and Optimism's includes \num{1993397} nodes and \num{43829952} edges. Tornado Cash's network has \num{34536} nodes and \num{272973} edges, while Uniswap's network includes \num{1355596} nodes and \num{4651837} edges.

Since our primary goal is to analyze where airdrop recipients sent their tokens in their first post-airdrop transfer, we restrict our analysis to a 1-hop subgraph that captures only the initial outbound transfers from recipient addresses.

\begin{definition}[1-hop subgraph]
Let \( R \subseteq V \) denote the set of addresses that received airdropped tokens directly from the protocol within the initial distribution window. The \textit{1-hop subgraph} \( G_1 = (V_1, E_1) \) is defined as:
\[
V_1 = R \cup \{ v \in V \mid (r, v) \in E \text{ for some } r \in R \}, \quad 
E_1 = \{ (r, v) \in E \mid r \in R \}
\]
\end{definition}

\begin{figure*}[]
\centering
\begin{subfigure}{\twocolgrid}
  \centering
  \includegraphics[width=.7\twocolgrid]{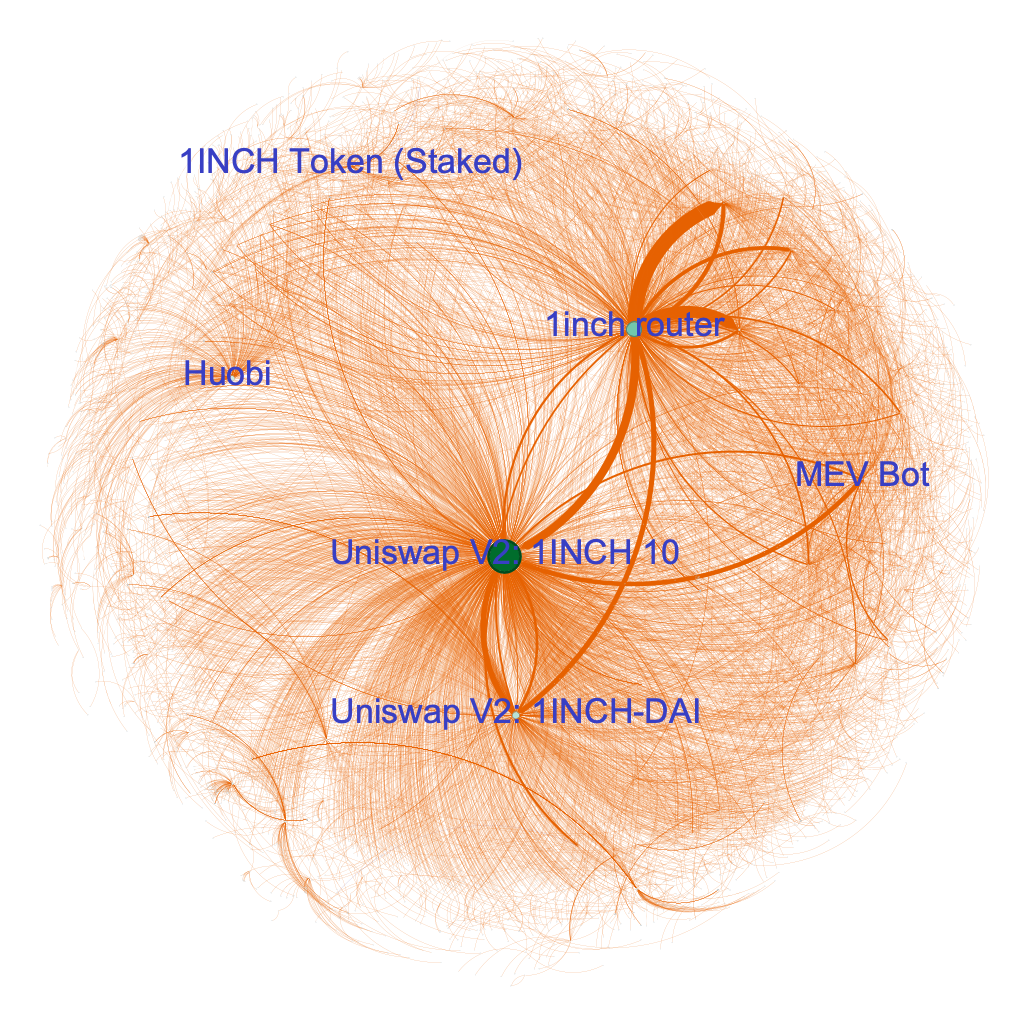}
  \caption{1inch.}
  \label{fig:airdrop-sale-networks-1inch}
\end{subfigure}
\begin{subfigure}{\twocolgrid}
  \centering
  \includegraphics[width=.7\twocolgrid]{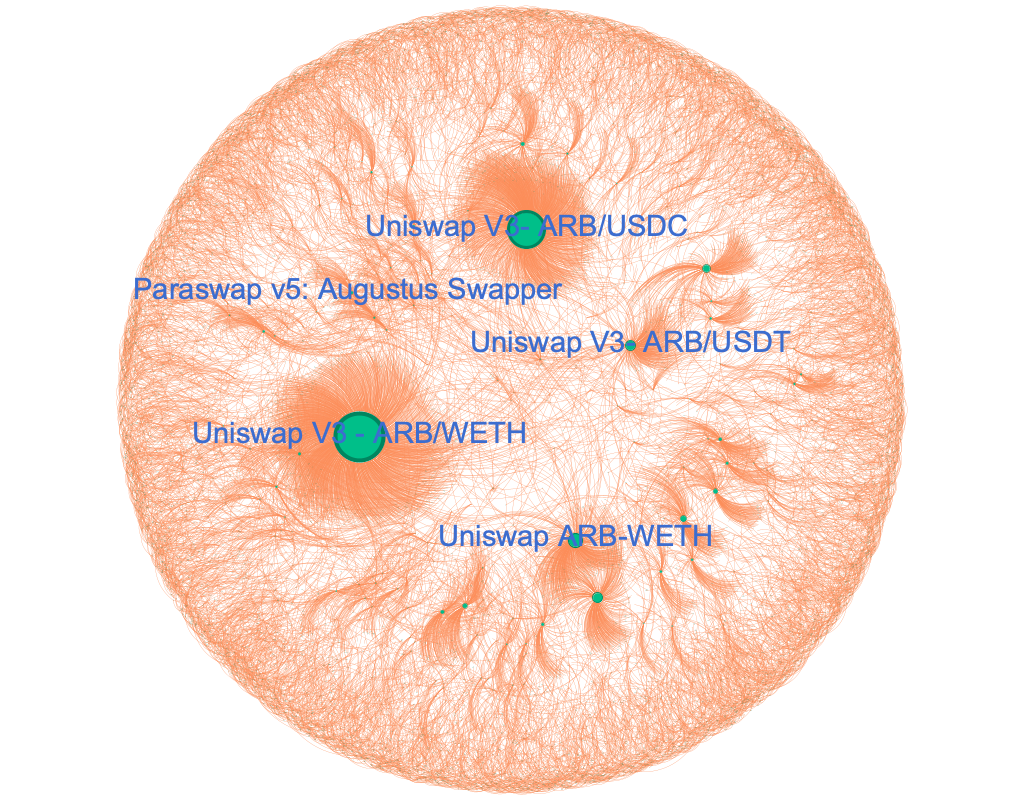}
  \caption{Arbitrum.}
  \label{fig:airdrop-sale-networks-arbitrum}
\end{subfigure}
\\
\begin{subfigure}{\twocolgrid}
  \centering
  \includegraphics[width=.7\twocolgrid]{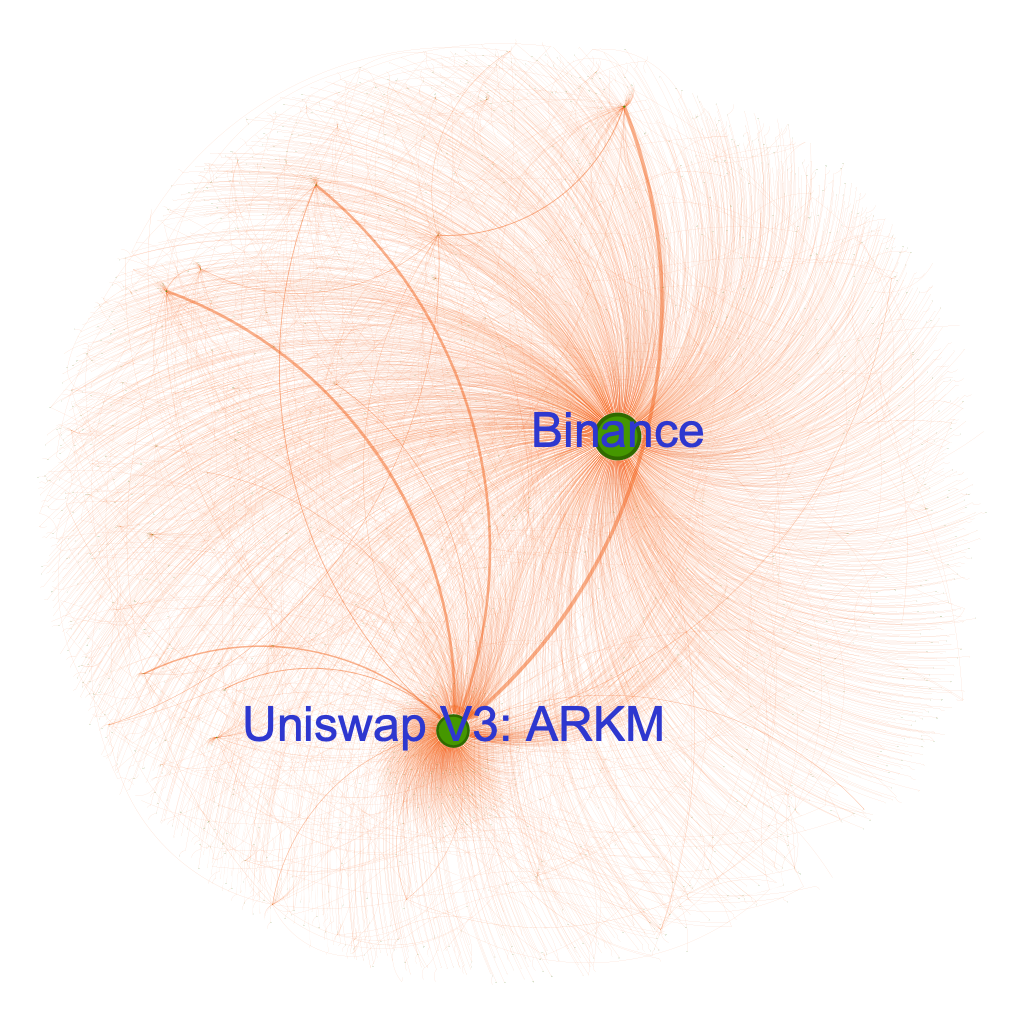}
  \caption{Arkham.}
  \label{fig:airdrop-sale-networks-arkham}
\end{subfigure}
\begin{subfigure}{\twocolgrid}
  \centering
  \includegraphics[width=.7\twocolgrid]{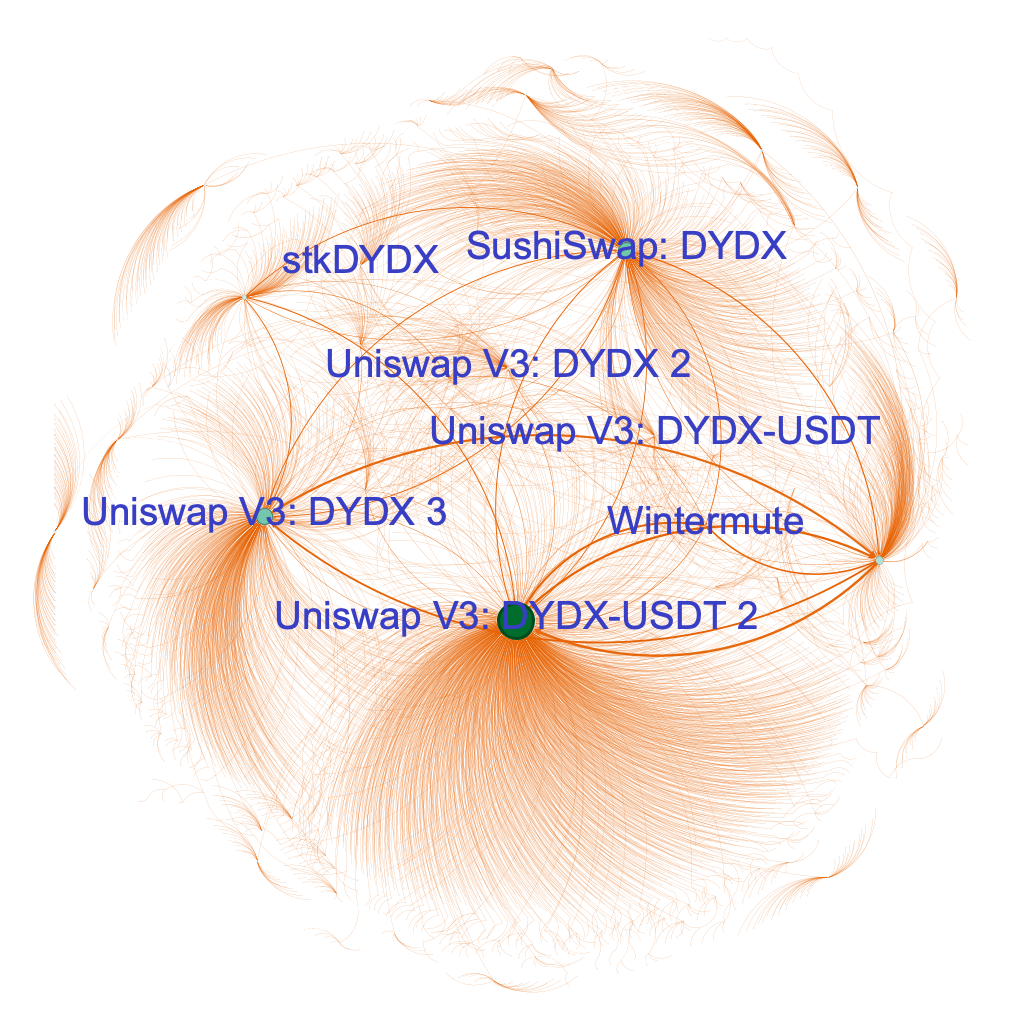}
  \caption{DYDX.}
  \label{fig:airdrop-sale-networks-dydx}
\end{subfigure}
\\
\begin{subfigure}{\twocolgrid}
  \centering
  \includegraphics[width=.7\twocolgrid]{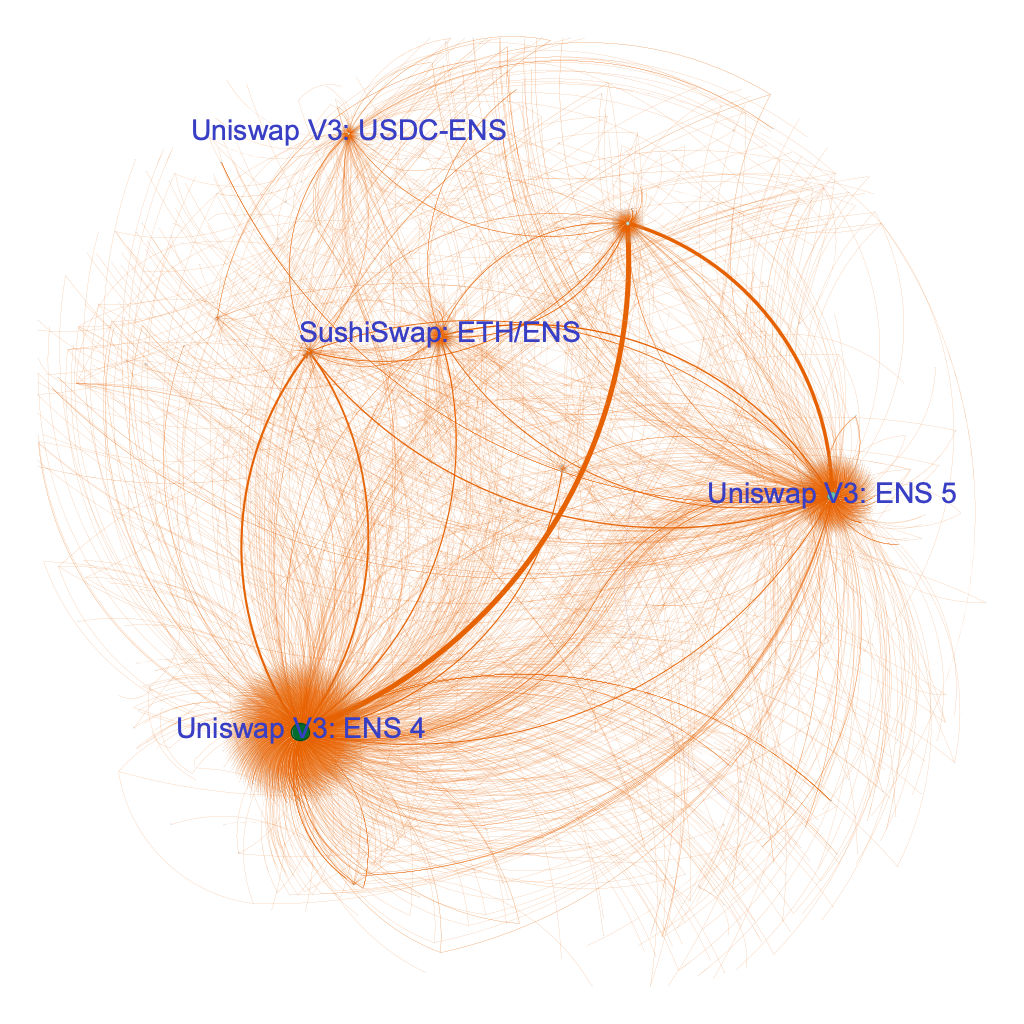}
  \caption{ENS.}
  \label{fig:airdrop-sale-networks-ens}
\end{subfigure}
\begin{subfigure}{\twocolgrid}
  \centering
  \includegraphics[width=.7\twocolgrid]{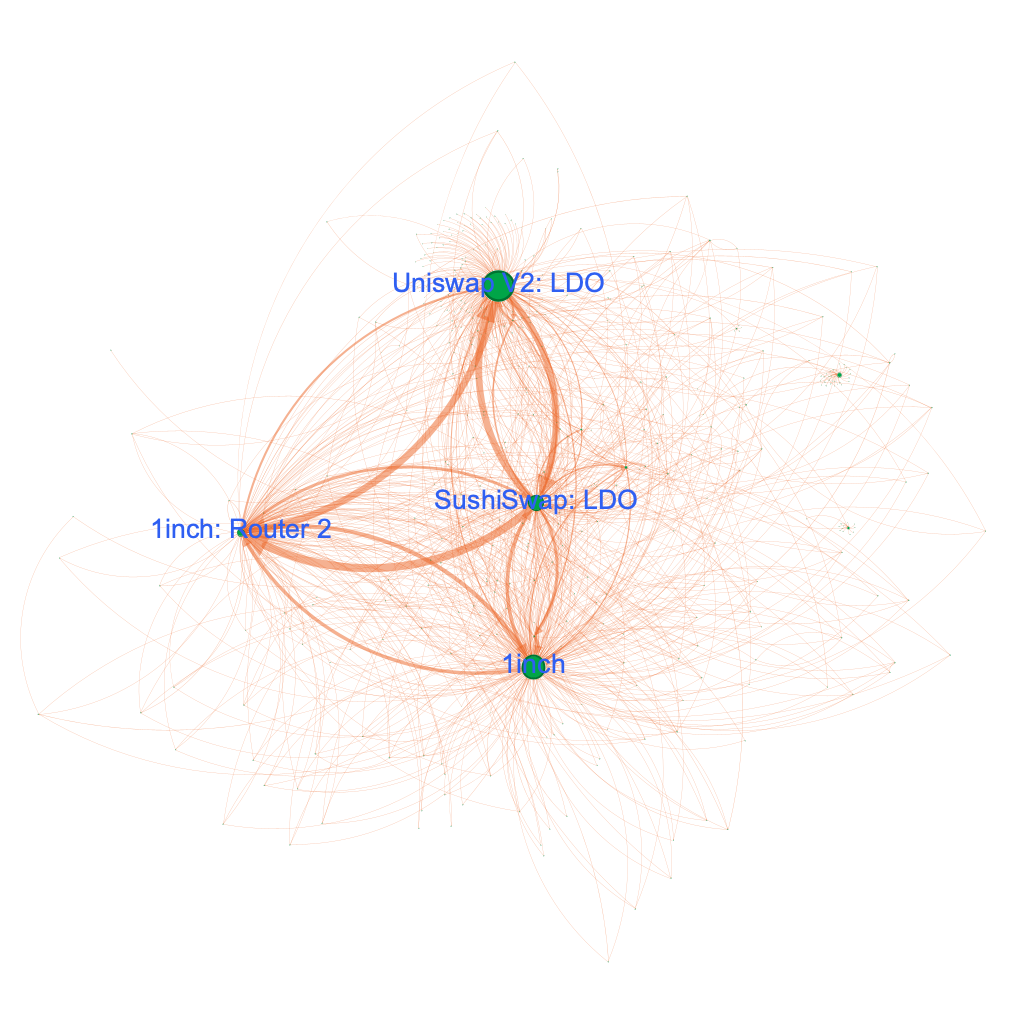}
  \caption{Lido.}
  \label{fig:airdrop-sale-networks-lido}
\end{subfigure}
\\
\begin{subfigure}{\twocolgrid}
  \centering
  \includegraphics[width=.7\twocolgrid]{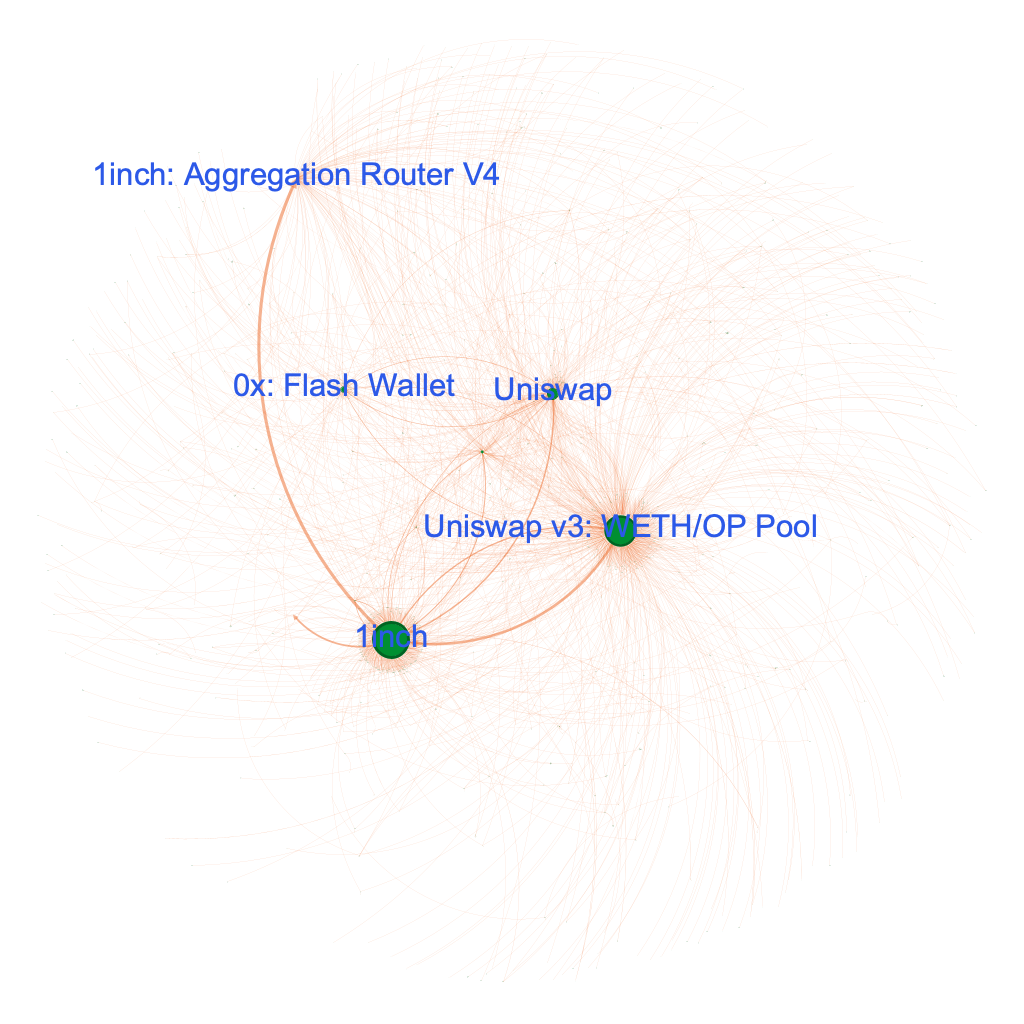}
  \caption{Optimism.}
  \label{fig:airdrop-sale-networks-optimism}
\end{subfigure}
\begin{subfigure}{\twocolgrid}
  \centering
  \includegraphics[width=.7\twocolgrid]{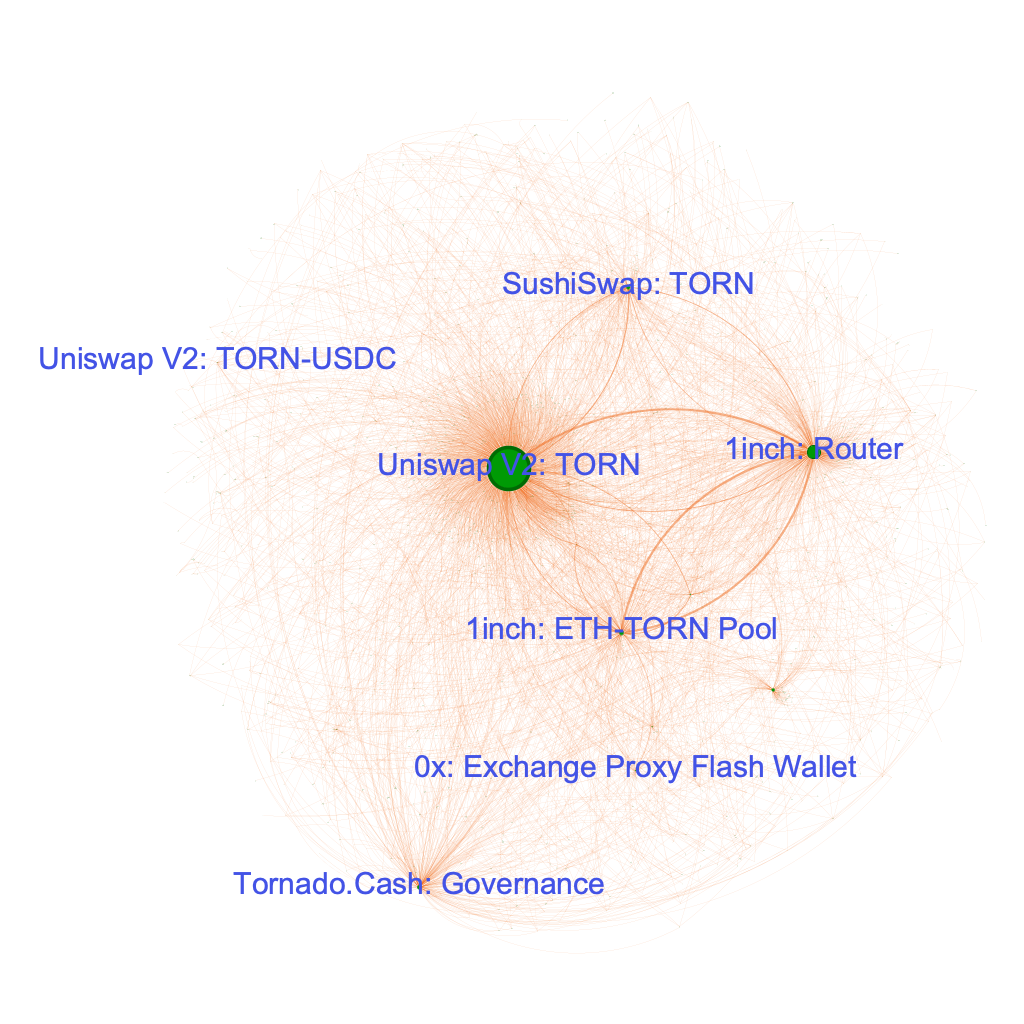}
  \caption{Tornado Cash.}
  \label{fig:airdrop-sale-networks-tornado}
\end{subfigure}
\\
\begin{subfigure}{\twocolgrid}
  \centering
  \includegraphics[width=.7\twocolgrid]{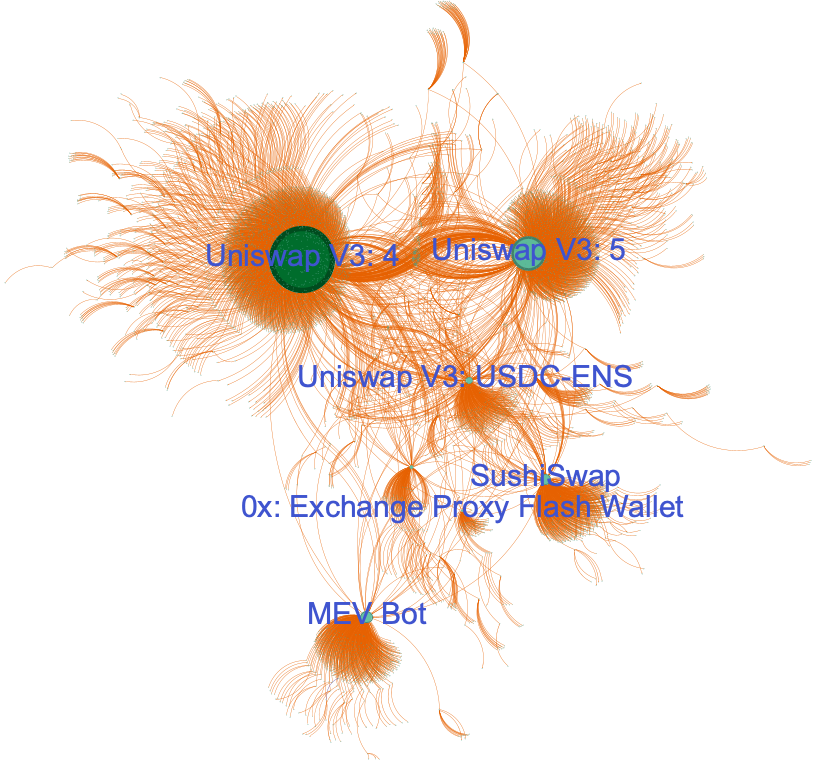}
  \caption{Uniswap.}
  \label{fig:airdrop-sale-networks-uniswap}
\end{subfigure}
\figcap{Network structure for selling airdropped tokens in nine airdrops.}
\label{fig:airdrop-sale-networks}
\end{figure*}

We visualize the largest weakly connected component of each \( G_1 \) using the Gephi tool, as shown in Figure~\ref{fig:airdrop-sale-networks}. To enrich the analysis, we manually annotate high in-degree nodes belonging to centralized and decentralized exchanges with labels from our dataset (\S\ref{sec:dataset}). Across all protocols, we find that the most frequent recipient of airdropped tokens is \stress{Uniswap}, followed by \stress{SushiSwap}. Among centralized exchanges, \stress{Binance} and \stress{Huobi} appear as the leading recipients.

\begin{figure}[t]
\centering
\begin{subfigure}{\twocolgrid}
  \includegraphics[width=\twocolgrid]{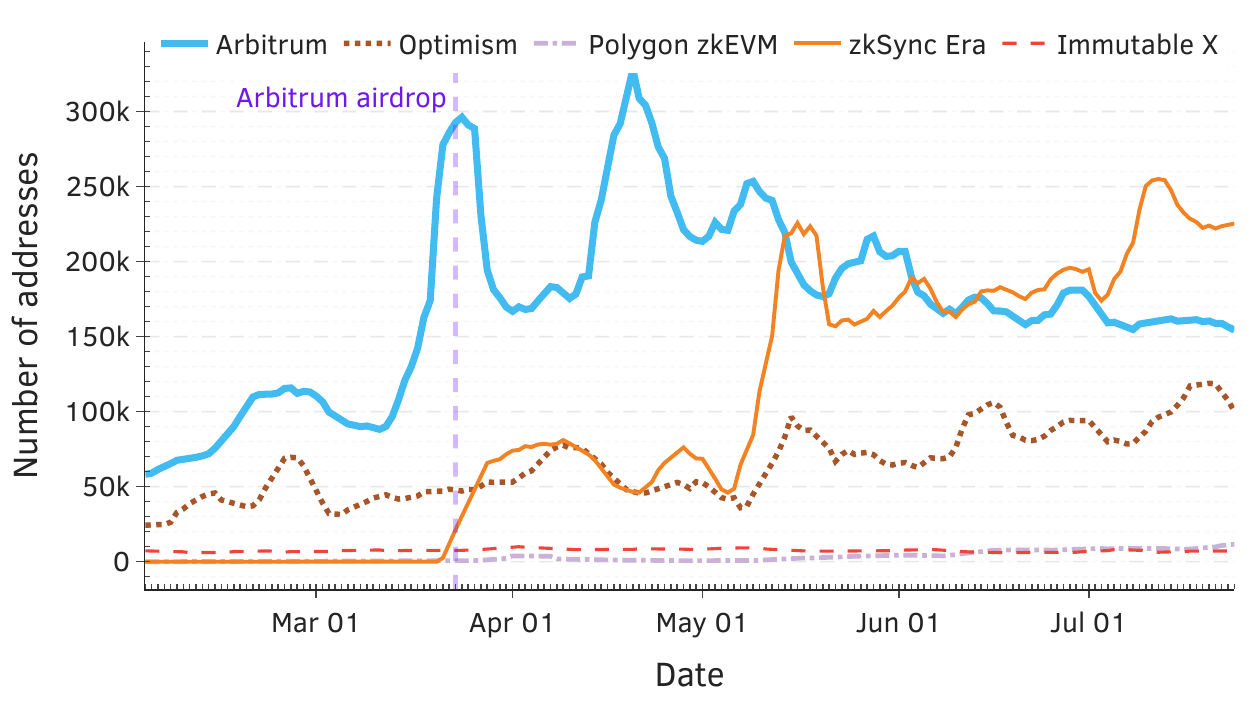}
  \caption{Absolute number.}
  \label{fig:addresses-absolute}
\end{subfigure}
\begin{subfigure}{\twocolgrid}
  \includegraphics[width=\twocolgrid]{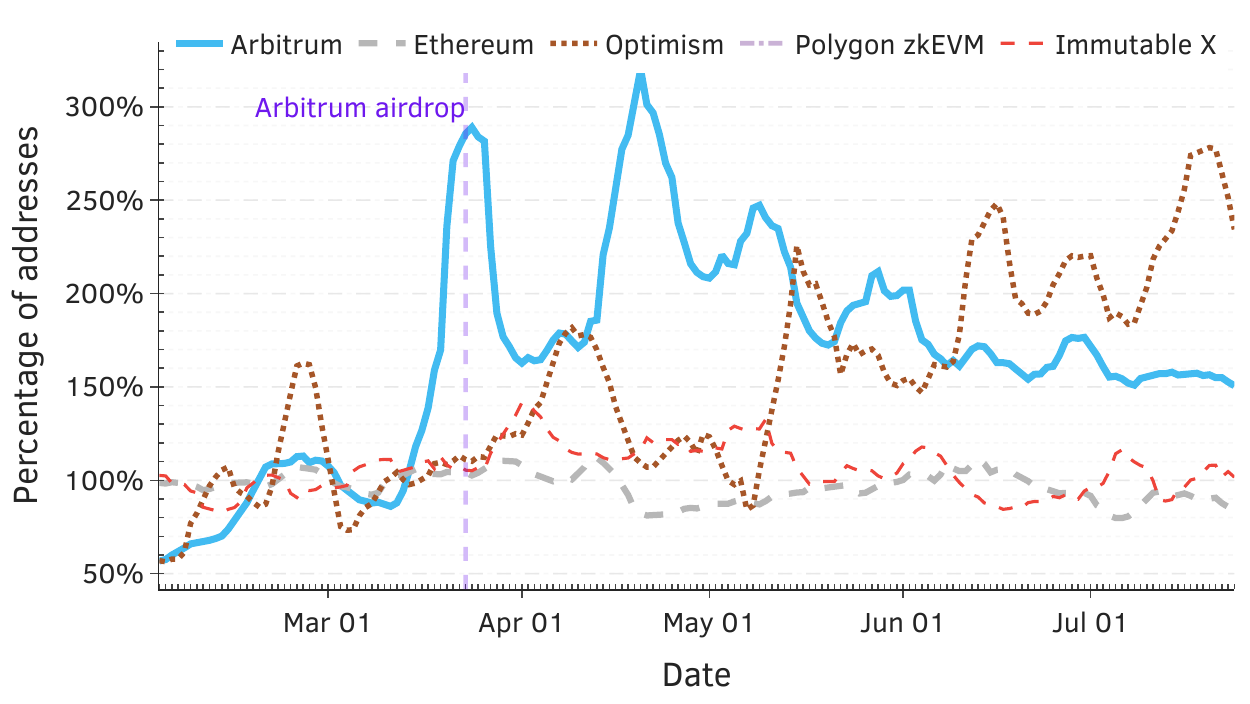}
  \caption{Relative daily active addresses.}
  \label{fig:addresses-relative}
\end{subfigure}
\caption{Unique daily active addresses per protocol: (a) Absolute number; (b) relative to the average for the 50 days before Arbitrum's airdrop.}
\label{fig:addresses}
\end{figure}

\begin{figure}[t]
\centering
\begin{subfigure}{\twocolgrid}
  \centering
  \includegraphics[width=\twocolgrid]{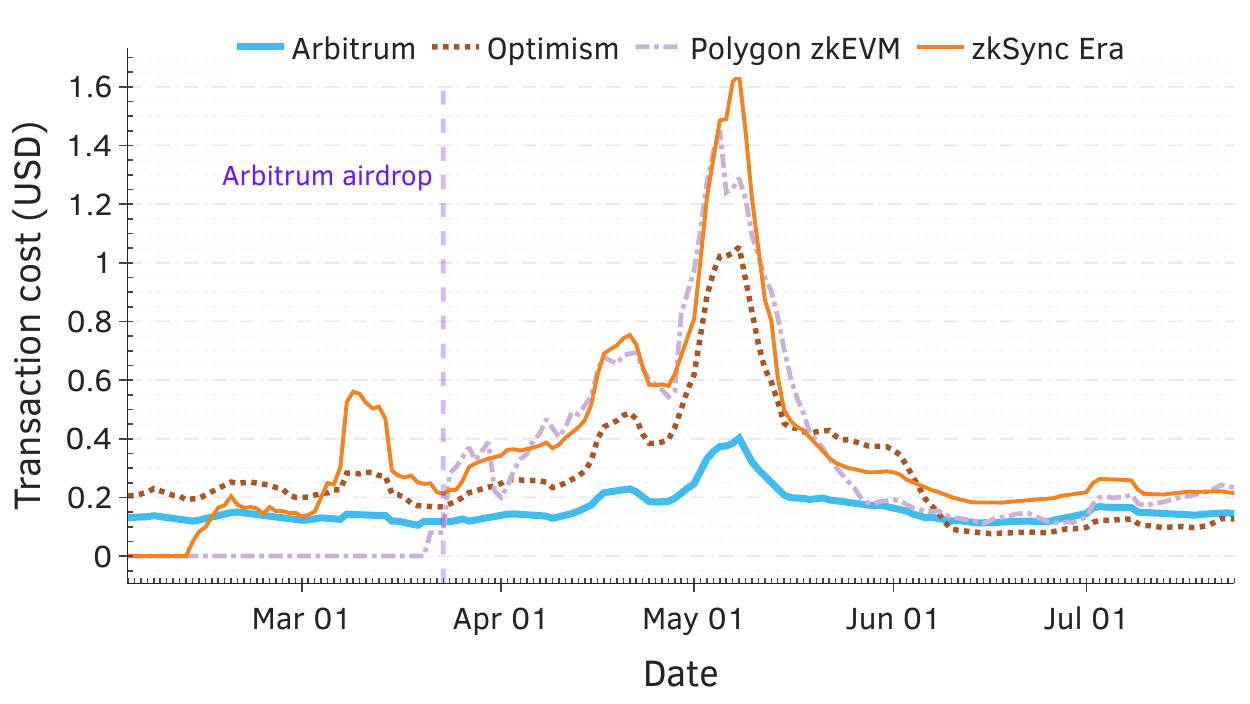}
  \caption{Daily median fees.}
  \label{fig:daily-fees-median}
\end{subfigure}
\begin{subfigure}{\twocolgrid}
 \centering
 \includegraphics[width=\columnwidth]{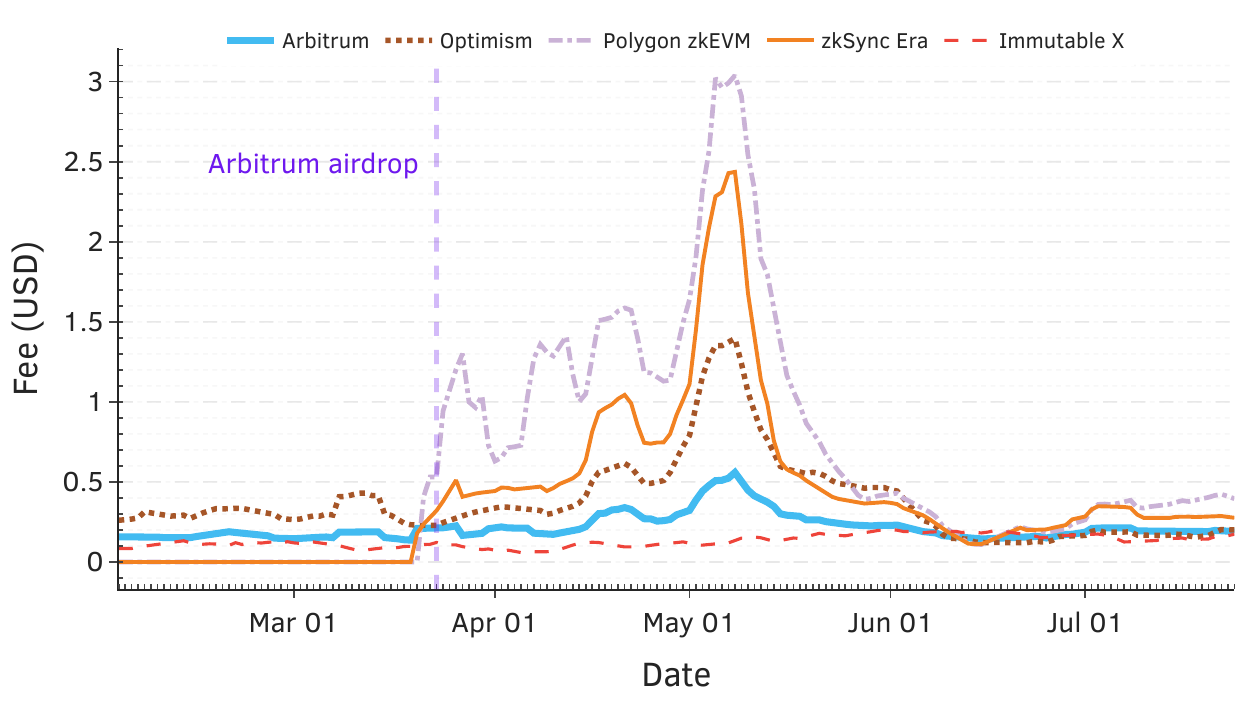}
 \caption{Daily average fees.}
 \label{fig:daily-fees-average}
\end{subfigure}
\figcap{Daily transaction fees in USD. (a) Median; and (b) Average.}
\label{fig:daily-fees-2}
\end{figure}

\section{Measuring the Airdrop Lift}
\label{sec:status-quo}

Empirical evidence suggests that some airdrops excel in attracting users in the short-term, at least superficially~\cite{fanAltruisticProfitorientedMaking2023}.
Although there is some preliminary data showing that airdrops fail in achieving the other goals, there is a dearth of solid research on the topic~\cite{fanAltruisticProfitorientedMaking2023,frowis2019operational}.
In this section, we examine this through the lens of some relevant metrics such as daily transaction count, daily active addresses, median transaction cost, \gls{TVL}, fees paid by users, and stablecoin market cap applied to the Arbitrum airdrop. We rely on data sourced from Growthepie~\cite{Growthepie}. For more details on this data refer to \S\ref{sec:dataset}.

\subsection{Unique Active Addresses}

\paraib{Others outperformed Arbitrum without an airdrop}
While the number of unique addresses on Arbitrum increased post-airdrop (March~22nd~2023~\cite{arbitrum-announcement}) and remained 50\% higher than pre-airdrop levels, other protocols achieved similar growth without an airdrop. For example, per Figure~\ref{fig:addresses-relative}, Optimism saw a larger increase in addresses in May 2023, likely due to the Bedrock launch~\cite{NijkerkCoindesk}. Similarly, ZKsync Era surpassed Arbitrum's address count two months after the airdrop, as shown in Figure~\ref{fig:addresses-absolute}.

\paraib{Fees may explain the closing gap between Arbitrum and Optimism}
Arbitrum has consistently led Optimism in daily active addresses. However, Figure~\ref{fig:DailyActiveAddressesRatio} in \S\ref{sec:airdrop_metrics} shows this gap narrowing. Before the airdrop, Arbitrum had 2.6 times more active addresses than Optimism, but in the last 50 days, this has decreased to 1.83 times. Optimism's lower median transaction fees since June may partly explain this trend (Figure~\ref{fig:daily-fees-median}), with the average fee gap between the two protocols being even smaller (Figure~\ref{fig:daily-fees-average}).

\paraib{Unique address counts oscillates over time}
The number of unique addresses shows oscillatory behavior, as seen in Figure~\ref{fig:addresses-relative}, with rapid rises, peaks, and drops. Notably, Optimism and Arbitrum display inverted phases in relative address counts, possibly due to users switching between protocols as fees rise. However, Figure~\ref{fig:daily-fees-median} shows no such pattern in median transaction fees, with Arbitrum consistently having lower fees in mid-May 2023. However, \stress{unique address counts metric can be gameable.} Per our analysis, Arbitrum's airdrop did not lead to long-lasting user involvement, as the unique address metric can be gamed. Users can create multiple addresses (Sybil attacks) to exploit airdrop limits. This makes the metric unreliable for measuring genuine activity, as large fluctuations may result from such behavior. Furthermore, the accessibility of airdrop farming software facilitates the automated execution of such activities~\cite{Combine,NFTCopilot,FarmerFriends,SybilSamurai}. Thus, alternative, more Sybil-resistant metrics should be considered to gauge authentic user involvement, such as graph network analysis~\cite{fanAltruisticProfitorientedMaking2023,liu2022fighting} combined with machine learning techniques~\cite{Hu@WWW23,moser2022resurrecting,victor2020address}.

\subsection{Transaction-related Metrics}
Transaction-related metrics provide a useful proxy for measuring ``real'' economic activity, as users must pay a fee to send transactions~\cite{gafni2022greedy}, excluding cases with retroactive rebates or protocols operating at a loss~\cite{Optimism-Airdrop-2}.

\begin{figure}[t]
\centering
\begin{subfigure}{\twocolgrid}
  \centering
  \includegraphics[width=\twocolgrid]{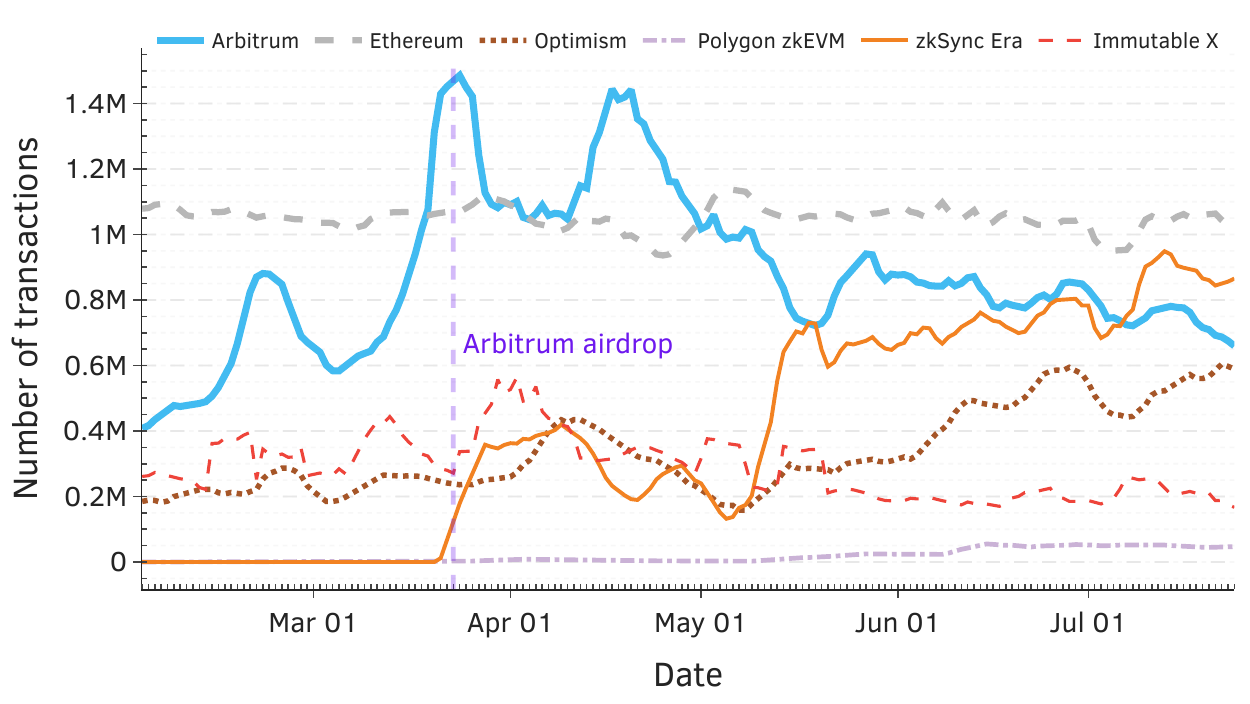}
  \caption{Absolute daily transactions.}
  \label{fig:daily-txs-per-platform-absolute}
\end{subfigure}
\begin{subfigure}{\twocolgrid}
  \centering
  \includegraphics[width=\twocolgrid]{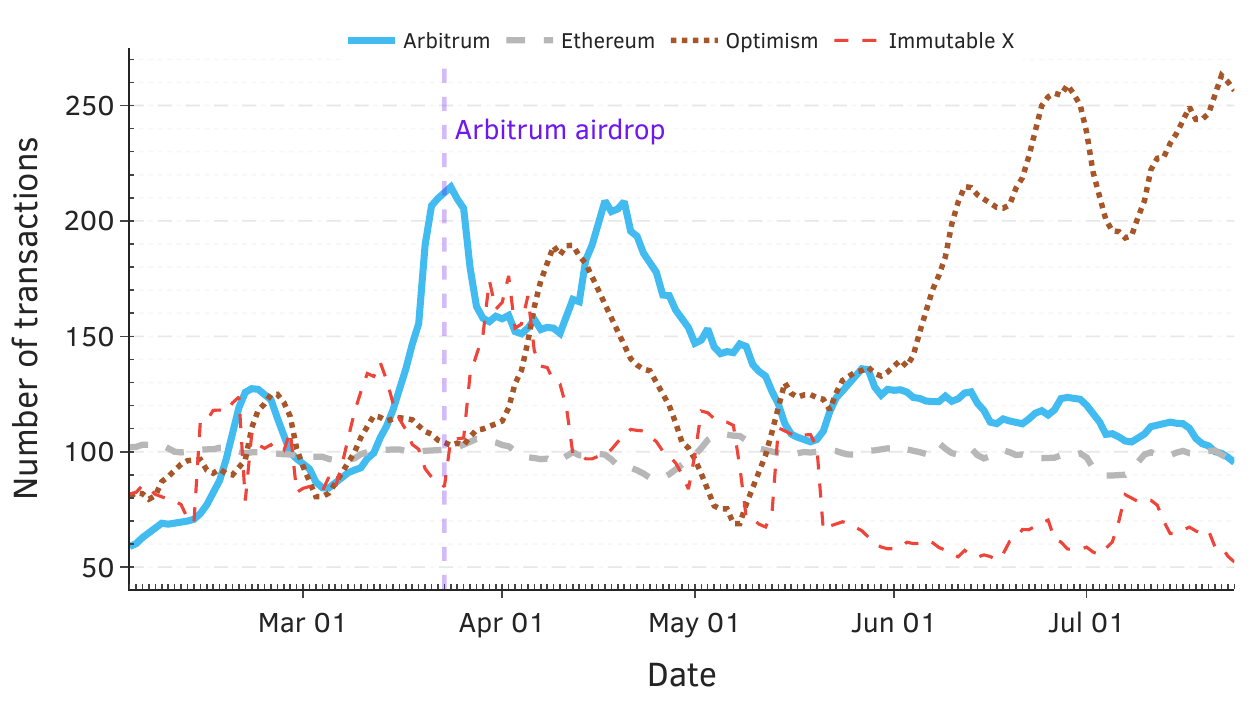}
  \caption{Relative daily transactions.}
 \label{fig:daily-txs-per-platform-relative}
\end{subfigure}
\caption{Daily number of transactions per protocol. (a) Absolute number; (b) relative to the average for the 50 days before Arbitrum's airdrop.}
\label{fig:daily-fees}
\end{figure}

\paraib{The Arbitrum-Optimism gap narrows when considering transactions}
Notably, the transaction count gap between Arbitrum and Optimism had nearly closed by the end of July (see Figure~\ref{fig:daily-txs-per-platform-absolute}). Additionally, Figure~\ref{fig:daily-txs-per-platform-relative} shows that Immutable X's daily transaction count has almost halved since Arbitrum's airdrop, whereas its unique address count remained relatively stable. This suggests that while address numbers held steady, user engagement with Immutable X has decreased.

\begin{figure}[t]
    \centering
\begin{subfigure}{\twocolgrid}
    %    \centering
        \includegraphics[width=\twocolgrid]{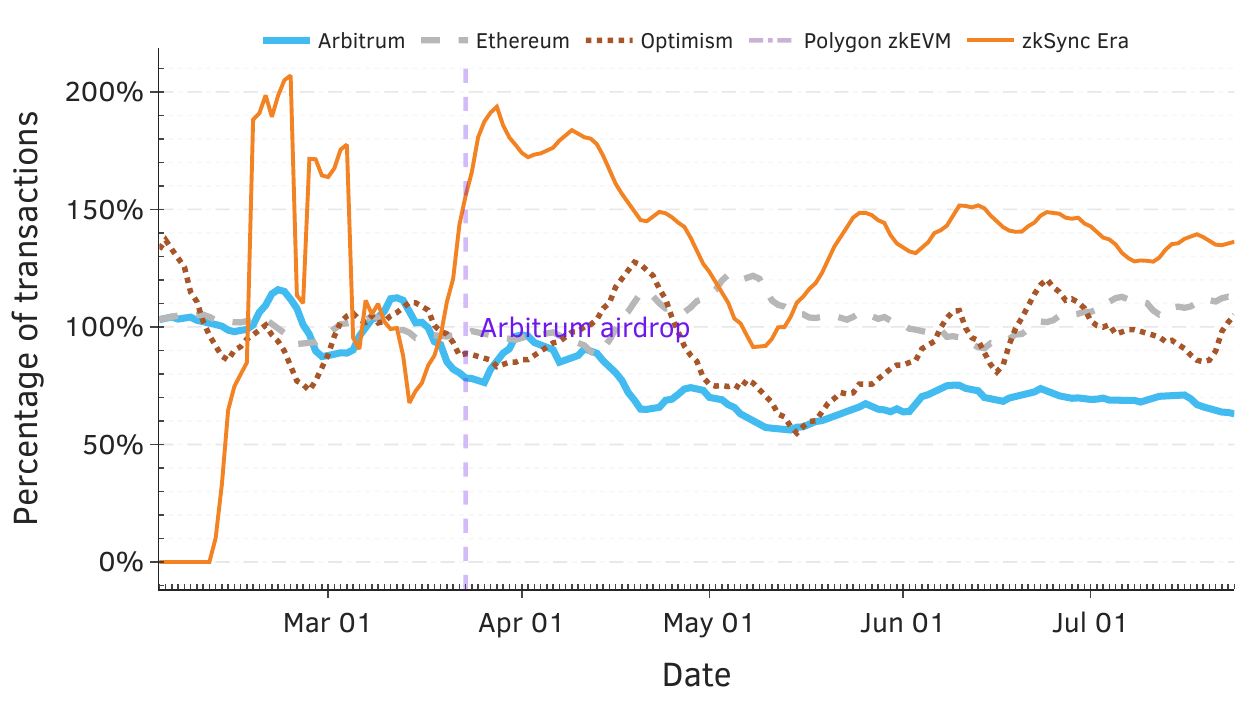}
        \caption{Daily transactions per address.}
        \label{fig:post-airdrop-relative}
\end{subfigure}
\begin{subfigure}{\twocolgrid}
        \centering
        \includegraphics[width=\twocolgrid]{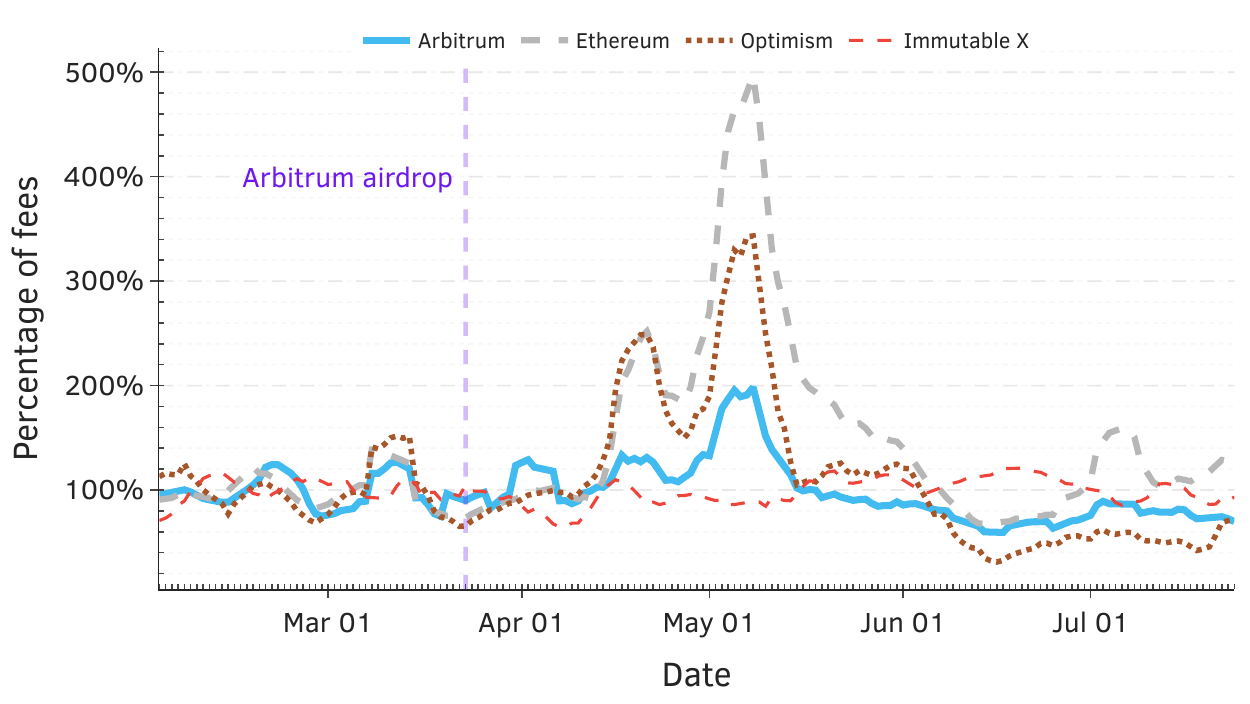}
        \caption{Average fees spent per address.}
        \label{fig:relative-fee-per-address}
\end{subfigure}
\caption{Daily transactions per address (a) and daily average fees per address (b), relative to the average before Arbitrum's airdrop.}
\label{fig:post-airdrop}
\end{figure}

\paraib{Arbitrum's per-address transaction count declined post-airdrop}
To assess how Arbitrum's airdrop impacted user engagement in other protocols, Figure~\ref{fig:post-airdrop-relative} shows the relative average daily transaction count per unique address. Since the airdrop, Arbitrum's per-user transaction count has dropped to less than 75\% of its pre-airdrop level. However, since \stress{transaction counts can be misleading without considering fees}, high transaction counts alone may not indicate genuine engagement. In this regard, some protocols require users to perform multiple transactions to qualify for airdrops, leading to inflated activity when fees are low. This aligns with Goodhart's law~\cite{manheim2019categorizing}, which states that \stress{``when a measure becomes a target, it ceases to be a good measure''}.

\paraib{Since June, average transaction fees are similar for all protocols}
Good metrics should reflect a user's commitment, and transaction fees can serve as a proxy for this since it measures how much a user is willing to expend for interacting with the protocol. 
Since June, the average fee per transaction and per unique address have been similar across protocols. Comparing average and median fees (Figure~\ref{fig:daily-fees-2}) shows that the median fee may be more useful in understanding shifts in user behavior.
Another useful metric is the relative average fees \emph{per address} compared to the 50 days before Arbitrum's airdrop, as shown in Figure~\ref{fig:relative-fee-per-address}. It shows that Arbitrum's user engagement was not significantly impacted by the airdrop and generally follows similar patterns to other protocols.

\begin{figure}[t]
    \centering
\begin{subfigure}{\twocolgrid}
    %    \centering
        \includegraphics[width=\twocolgrid]{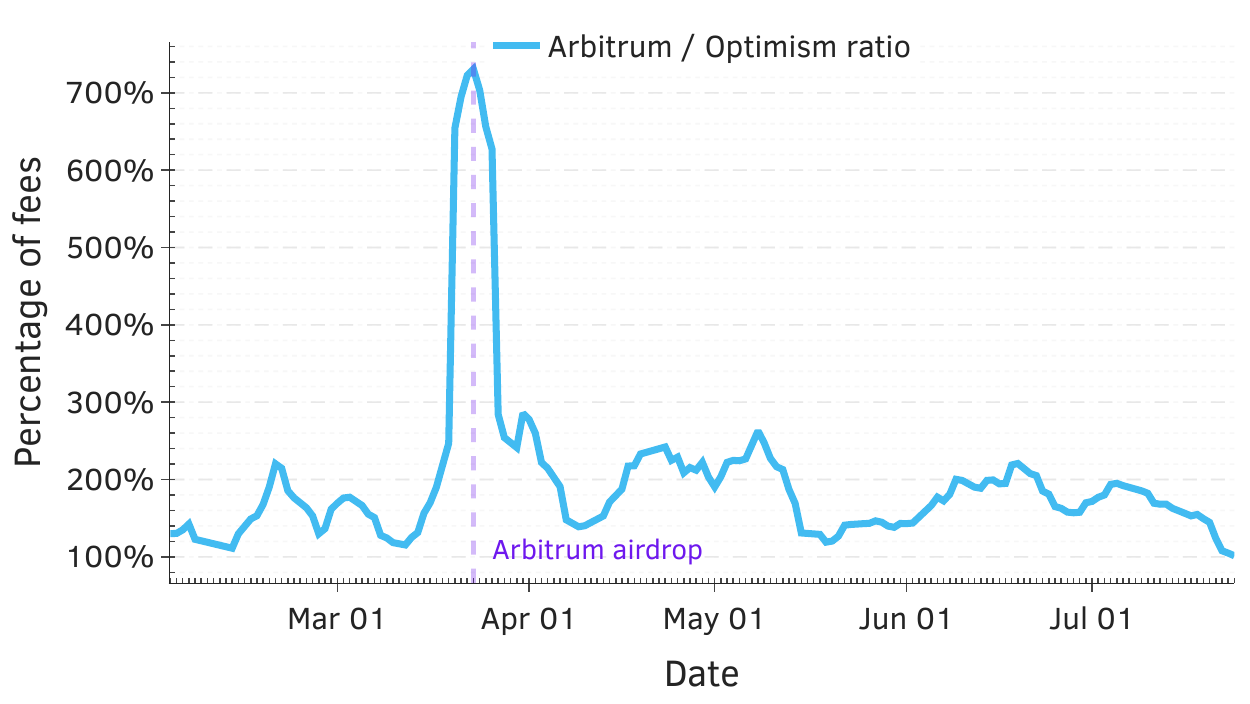}
        \caption{Ratio of daily transaction fees.}
        \label{fig:ratio-arb-opt}
\end{subfigure}
\begin{subfigure}{\twocolgrid}
        \centering
        \includegraphics[width=\twocolgrid]{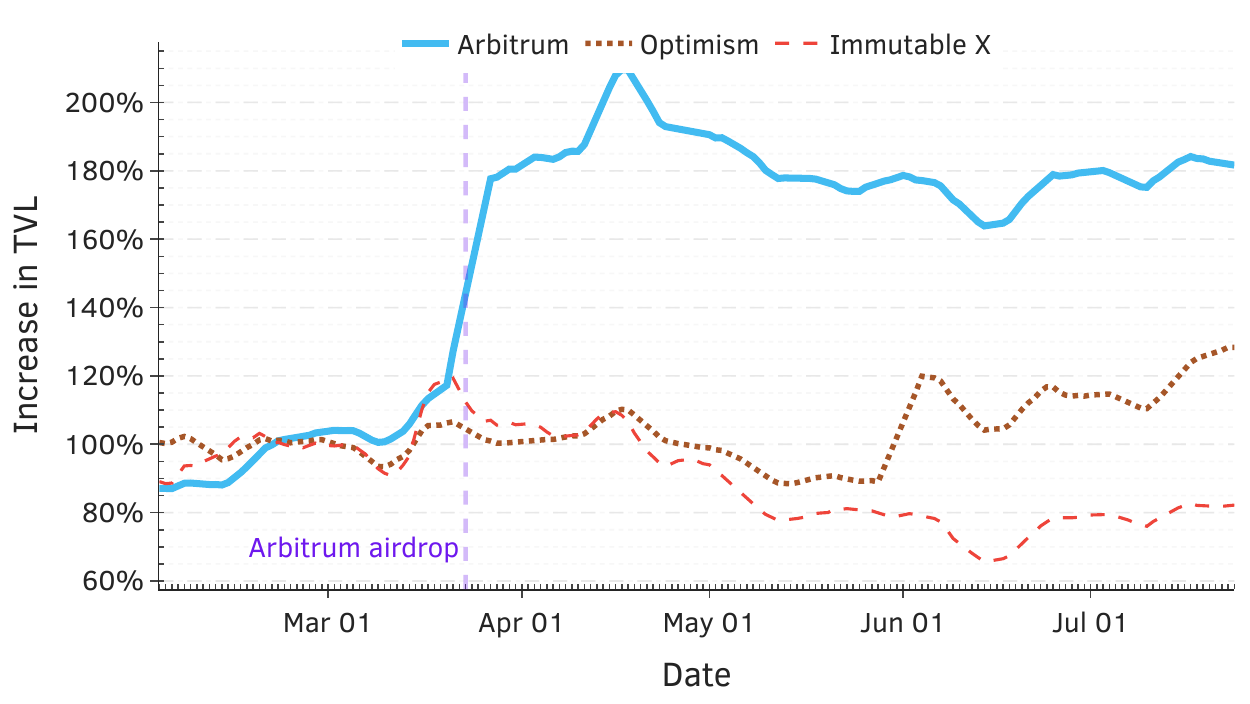}
        \caption{TVL growth.}
        \label{fig:tvl-growth}
\end{subfigure}
\caption{Comparison of the ratio of daily transaction fees between Arbitrum and Optimism (a); and the growth in Total Value Locked (TVL) relative to the average of the 50 days before Arbitrum's airdrop (b).}
\label{fig:ratio-fees-and-tvl}
\end{figure}

\paraib{Arbitrum's total daily fees spiked around the time of the airdrop}
Arbitrum's airdrop did not provide a significant long-term advantage over its competitors in terms of income from transaction fees as shown in Figure~\ref{fig:fees-before-after} in \S\ref{sec:airdrop_metrics}. While Arbitrum experienced a peak in fees on the day of the airdrop, this increase was short-lived.
Indeed, per Figure~\ref{fig:ratio-arb-opt}, in the~50 days before the airdrop, Arbitrum earned on average~1.96 times more from transaction fees per day, while in the last~50 days of our data set this has shrunk to~1.74.

\subsection{Total Value Locked (TVL)}
The \gls{TVL} of a protocol is a metric that measures the combined value of all assets stored within that protocol.

\paraib{Arbitrum's airdrop had a lasting impact on its TVL}
Among all examined metrics, TVL is the only one that showed a long-lasting improvement following an airdrop: \stress{Arbitrum's TVL increased by over~50\% immediately afterwards, and has not dropped significantly since}, as shown in Figure~\ref{fig:tvl-growth}.
It is perhaps surprising, as Arbitrum's airdrop distribution strategy considered only user activity performed before February 6\tsup{th}, 2023~\cite{Arbitrum-allocation}.

\section{Common Airdrop Design Challenges} 
\label{sec:pitfalls}

Airdrops, much like traditional loyalty programs—such as newcomer bonuses provided by banks and credit card companies—face several common design challenges. However, the unique context of blockchain technology and the specific mechanics employed by most airdrops can amplify these challenges or even introduce new ones. In this section, we explore three of these challenges.

\subsection{Airdrop Farmers}\label{subsec:airdrop-farmers}
These are sophisticated users who employ elaborate tactics to maximize the amounts of airdrop tokens they receive.
Blockchain protocols have adopted various measures to mitigate airdrop farmers gaming the system.
A popular approach is to limit the amount of rewards that a single user can receive.

Thus, protocols resort to \gls{PoH} services, such as Gitcoin passport~\cite{Gitcoin-passport}. 
These usually assign a numeric score to users based on a certain metric, where a higher score means that there is more certainty that the corresponding user is indeed human.
Gitcoin passport's metric is based on a list of tasks, such as connecting a user's social media accounts, or being in possession of some amount of ETH~\cite{Gitcoin-passport}.
Such methods can be augmented by analyzing on-chain data to detect Sybils and exclude them from the airdrop~\cite{Arbitrum-airdrop}, but this may lead to false negatives~\cite{Arbitrum-analysis}. There are some recent examples of protocols like Celestia using Sybil resistance solutions~\cite{Celestia-airdrop}.

There are other mitigation techniques, such as requiring users to perform tasks, ranging from sending a certain type of transactions~\cite{HAPI-Protocol}, to sharing posts on social media~\cite{BulletLabs}. These tasks sometimes appear arbitrary, leading to frustration among users and they are easy to automate~\cite{WeAreCrypto} and cheaply fooled, even more-so when protocols offer retroactive rebates on transaction fees~\cite{Optimism-Airdrop-2}.
Furthermore, the reliance of many protocols on a limited number of \gls{PoH} services means that a one-time investment by an airdrop farmer can result in large profits from multiple airdrops, even with biometric identification, full Sybil resistance is not guaranteed~\cite{TrustaLabs}.
A different approach used by protocols is to announce airdrops that retroactively reward users that were active before the announcement.
Nevertheless, farmers can prepare in advance by interacting with these protocols even when no airdrop has officially been announced~\cite{TheBlock}, as was the case in DYDX's airdrop~\cite{Edwards@Finder}.

The phenomena of reward farming is not limited to crypto-related airdrops, and similar ones can be found in ``traditional'' loyalty programs.
In particular, the practice of credit card churning, where users apply for credit cards for the sole purpose of receiving the rewards bestowed upon newcomers, only to cancel the cards once the rewards were received. 
Given that similar problems exist even in traditional contexts, where users can be easily identified and penalized, it seems that reward farming cannot be solved easily.

\subsection{Decentralized Governance Threats}
Some protocols distribute \textit{governance tokens} through airdrops as a means to decentralize their governance processes. However, \textit{distributing governance tokens can be risky}. These tokens allow holders to participate in the protocol's governance by voting on key decisions. Typically, these tokens can also be exchanged for other tokens, potentially giving them monetary value~\cite{dotan2023vulnerable}. This can lead to more farmers farming these tokens.

Empirical evidence suggests that airdropping governance tokens may be more effective than those of non-governance tokens. In a recent analysis, Makridis \textit{et al.}~\cite{makridis2023rise} found that airdropped governance tokens experienced a growth rate in market capitalization up to \num{14.99}\% higher compared to governance tokens that were not airdropped. However, the authors also note that this effect is statistically insignificant when using common benchmarks.

Despite these potential benefits, airdropping governance tokens poses significant risks if not done well. It can concentrate too much power in the hands of a few users, leading to an inequitable distribution of decision-making power within the system~\cite{feichtinger2023hidden,messias2023understanding,sharma2024unpacking}. Additionally, some recipients may not have the protocol's best interests in mind and could act in ways detrimental to the protocol's long-term success by voting on changes to the protocol that benefit themselves.

\subsection{Insider Trading}
This issue arises when individuals exploit privileged information for financial gain at the expense of other protocol users. Such practices are widely recognized as violations of securities laws in traditional financial markets~\cite{bainbridge2013overview} and often provoke negative reactions from the blockchain community~\cite{felez2022insider,Nelson@Coindesk,Solimano@AltLayer-TheDefiant}.

When someone within a protocol uses privileged information to increase their profits, it can trigger a community backlash. Insiders may have advanced knowledge of the metrics used to determine eligibility and the reward for each address, and may use it to their advantage. For example, there were claims that a head of growth at AltLayer might have used insider information to make a \num{200}K USD profit from their airdrop, although later dismissed as a mere coincidence~\cite{Solimano@AltLayer-TheDefiant}. However, such incidents can undermine the trust users place in these protocols.

This issue also raises concerns about fairness, as some users have access to superior and more accurate information than others. Identifying these inside traders is a challenging problem, making it crucial for protocols to provide extensive information to their users. In addition, incentives for blockchain data analytics companies and research groups to conduct post-airdrop data audits can help identify insider traders by analyzing address details and transfer patterns~\cite{felez2022insider}. For this to happen, data availability is key. Therefore, protocols should ensure transparency and encourage thorough analysis to maintain integrity and fairness within the blockchain community.

\section{Design Guidelines}
\label{sec:lessons-learned}

The design challenges outlined earlier, while sobering, may inform future airdrop designers, and can shed light on the path to potential success.

\subsection{Alternative Incentives for Sustaining User Engagement}

The long-term benefits of an airdrop to a protocol can be difficult to measure, as the potential advantages may be indirect and hard to quantify, while the expenses are often immediate and irrevocable. 
Additionally, some costs and effects, such as those arising from distributing governance tokens, may be unpredictable.
Therefore, instead of using airdrops, communities could consider alternative measures with a more predictable relationship between costs and benefits. One straightforward alternative is for the community to vote to programmatically reward loyal users with discounts on future interactions. In the context of Layer 2 (L2) solutions, these discounts could be applied to transaction fees. This approach encourages users to engage with the protocol again to benefit from the incentive, thereby fostering continued user participation. Additionally, this incentive mechanism is somewhat more resistant to airdrop farming, as the discounts have no intrinsic value outside the protocol, and the protocol's costs are directed solely toward users who actively use the system.

However, careful design of the discount mechanism is essential, including determining eligibility criteria for users and setting appropriate discount levels. Moreover, it is not immediately clear whether discounts can attract users as effectively as the more immediate and tangible rewards offered by airdrops.
Another option is to perform multiple airdrops over an extended period instead of a single event. Although this approach remains susceptible to some pitfalls of standard one-shot airdrops, it may help ensure long-term community engagement and prevent the decline in user adoption that some protocols experience immediately after an airdrop.
A more innovative approach was adopted by Blast, which introduced a point-based reward program~\cite{Melo@Dappradar}. Under this program, users earn rewards based on points accumulated through various activities, such as bridging tokens to the protocol (i.e., transferring funds from another protocol) and participating in a referral program that increases points for bringing in more users. Notably, Blast received \$1.1 billion in deposits even before its official launch~\cite{Knight@Coindesk}. This method provides a measurable metric for users' contributions to the protocol, structured around a referral-program model.

Furthermore, the incorporation of innovative distribution mechanisms plays a crucial role in mitigating the presence of sybil attacks within whitelisted addresses eligible for an airdrop. For example, Celestia has proposed a unique design that uses GitHub commits as a proxy to assess a user's contribution to the blockchain ecosystem~\cite{Celestia-airdrop-distribution}. However, a potential concern arises with the possibility of users or farmers generating counterfeit activities on GitHub to exploit other protocols' airdrops adopting a similar strategy to Celestia's. Farmers, therefore, might expect that new protocols use similar selection criteria than past airdrops. To counter this, \stress{protocols could focus on metrics that are resistant to programmable manipulation}, thereby adding a layer of difficulty or cost to the creation of automated user accounts.

\subsection{Targeting Well-Known Reputable Entities}

Instead of rewarding pseudonymous users, protocols can target developers and projects that are building relevant applications. For example, in Arbitrum's airdrop, 1.13\% of the distributed tokens were allocated to DAOs projects~\cite{Arbitrum-allocation}. Arbitrum also offers additional incentives beyond its airdrop for specific groups, such as university students~\cite{Arbitrum-embassador-program-university} and members of the technical community~\cite{Arbitrum-embassador-program} who wish to participate in research and develop tools relevant to the protocol.

Another approach was implemented by Optimism, which allocated some of its revenue to retroactively fund successful projects, essentially bringing the concept of a startup to the blockchain world~\cite{Optimism-goods-funding}.
\stress{Prioritizing established and reputable entities, including those building on the protocol, research groups, tech communities, and students, can promote sustained engagement}. By funding these entities, protocols may attract value-driven users and foster long-term involvement.

\subsection{Proactive Oversight and Community Engagement}

During the airdrop process, continuous monitoring and analysis of the protocol's data are crucial to prevent exploitation. A notable example is provided by Linea, whose team identified a vulnerability that allowed users to manipulate their incentive mechanism~\cite{Linea@Twitter}. This prompt discovery averted cheaters from claiming over one-third of the \glspl{NFT} offered as incentives~\cite{Rusty378@Dune}. %\footnote{A dashboard is available on Dune at \href{https://dune.com/rusty378/guildxyz-pins}{https://dune.com/rusty378/guildxyz-pins}.}.

In addition to technical oversight, engaging the community by maintaining open communication channels and offering bug bounties encourages the disclosure of vulnerabilities, whether exploited or not~\cite{CheatSheets}. For instance, an AzukiDAO community member disclosed a vulnerability, allowing the protocol to quickly address it~\cite{MetaSleuth}. Oversight should also extend beyond on-chain data. For example, social media is frequently exploited by scammers promoting fake airdrops, luring users into connecting their wallets to fraudulent websites that drain their funds. Even protocols not planning an airdrop may be targeted by such schemes.

Active community involvement in technical discussions is also advantageous. For example, ZKsync Era's \gls{NFT} drop was retrospectively analyzed by cygaar~\cite{cygaar-zksync}, who identified potential cost-saving improvements. \stress{Maintaining transparency and providing the community with insights into the protocol's inner workings fosters trust}. When technical issues arise, users are more likely to respond with understanding when they are well-informed.

\subsection{Rewards Should Scale with Costs}

The effects of Goodhart's law~\cite{manheim2019categorizing} are evident in many past airdrops. For example, airdrops often aim to incentivize user engagement by publicly announcing that eligibility for rewards is based on the frequency of past interactions with the protocol. This approach, however, can be exploited, as users may engage in meaningless wash-trading transactions~\cite{victor2021detecting}, undermining the criteria as a true reflection of genuine engagement.

The problem is further exacerbated by the fact that action-based metrics used to determine eligibility often fail to consider the actual costs users incur for each action. For instance, when the number of transactions is the primary metric, low transaction fees enable airdrop farmers to cheaply meet the transaction volume required for rewards. A potential solution is the adoption of a reputation-based mechanism~\cite{ReputationMechanism@A16Z}, which could disincentivize artificial inflation of transaction volumes. However, the protocol must carefully define what constitutes ``user reputation'' and determine the appropriate metrics for evaluation. 

Conversely, high transaction fees can erode the value of the rewards users receive, reducing the attractiveness of the airdrop~\cite{Friend@crypto}. To address these issues, \stress{rewards should scale with the actual costs incurred by users, ensuring a fairer and more effective distribution of incentives}.

\section{Related Work} 
\label{sec:related-work}

Recent research on airdrops has primarily focused on post-factum analyses and guidance for designing effective airdrop campaigns.

\paraib{Airdrop Studies} 
Yaish and Livshits~\cite{yaish2024tierdrop} present a theoretical model for airdrops that accounts for both honest users and farmers, where the latter have lower airdrop eligibility costs and lower intrinsic utility from using the platform that issues the airdrop. In their analysis, they show that airdrops in which the issuer expends a non-zero fixed cost for each recipient, the threat posed by farmer sybil attacks lead to unbounded issuance costs. However, they show losses from farmers can be bounded by setting in advance the total quantity of airdrop tokens and splitting them equally among recipients. Moreover, by designing the airdrop mechanism correctly, farmers can be harnessed to foster network effects, thereby attracting honest users who may otherwise would prefer competing platforms.

Makridis et al.~\cite{makridis2023rise} explore the impact of governance token airdrops on \gls{DEX} growth. Analyzing 51 exchanges, they found that such airdrops significantly contribute to increased market capitalization and trading volume.
Lommers et al.~\cite{lommersDesigningAirdrops2023} provide a comprehensive overview of various airdrop types, such as plain vanilla, holder-based, and value-based models. Their work highlights how eligibility criteria, signaling, and implementation strategies influence the success of airdrop campaigns, offering practical recommendations for optimization.
Fan et al.~\cite{fanAltruisticProfitorientedMaking2023} present a case study of ParaSwap DEX, proposing a role taxonomy methodology based on user behavior and airdrop effectiveness. Their findings show that users receiving higher rewards are more likely to contribute positively to the community. They also identify arbitrage patterns and note the limitations of current methods for detecting airdrop hunters. On the other hand, Sybil detection techniques, such as graph network analysis~\cite{fanAltruisticProfitorientedMaking2023,liu2022fighting} and machine learning approaches~\cite{beres2020blockchain,Hu@WWW23,moser2022resurrecting,victor2020address,Zhou-ARTEMIS@WWW24}, have been proposed to address these issues.

Allen~\cite{allenCryptoAirdropsEvolutionary2023} conducts nine airdrop case studies (e.g., Optimism, Arbitrum, Blur) and offers insights into designs like task-based claiming. The study emphasizes the need for more dynamic designs with feedback loops and suggests that some projects may revert to simpler designs due to the complexity and cost of advanced mechanisms.
Allen et al.~\cite{allen2023why} investigate the motivations behind token airdrops, focusing on marketing and decentralization. While airdrops are often seen as marketing tools, the authors argue that this rationale is weak, as evidence of marketing-driven airdrop success is limited. Instead, decentralization and community-building are highlighted as the primary motivations.

\paraib{Technical Aspects of Airdrops} 
Frowis et al.~\cite{frowisOperationalCostEthereum2019a} identify the operational challenges and costs associated with large-scale airdrops on Ethereum. They suggest that while cost savings of up to 50\% are possible with specific smart contract optimizations, pull-based approaches can shift the cost burden to recipients. Overall, the total cost remains proportional to the number of recipients.

Wahby et al.~\cite{wahby2020airdrop} address the privacy issues in current airdrop mechanisms, which leak information about recipients. They propose a private airdrop scheme based on zero-knowledge proofs using RSA credentials, achieving privacy while maintaining computational efficiency. Their implementation significantly improves both signature generation and verification times.

\section{Conclusion}
\label{sec:conclusion}

This study identified key challenges in the implementation of airdrops and proposed strategies to enhance their effectiveness. By analyzing nine major airdrops, we observed a recurring pattern where recipients rapidly sold their tokens—with 65.75\% of Lido, 58.67\% of 1inch, and 48.21\% of Optimism tokens traded shortly after distribution, within just 1-hop. This pattern suggests that current airdrop mechanisms may struggle to maintain long-term engagement and attract valuable contributors.

For Arbitrum, we identified a temporary surge in daily fees during the airdrop, followed by a decline in transactions per address. Notably, other protocols without airdrops performed similarly in transaction volume, and since June 2023, transaction fees have stabilized across all observed protocols. This finding underscores that airdrops alone may not be sufficient to drive sustainable user growth.

We also highlighted several key challenges, including airdrop farming, governance token distribution, and insider trading. Addressing these issues can provide critical insights for designing more effective airdrop strategies. To that end, we propose the following guidelines:

\paraib{Implement Alternative Incentives} Protocols should consider offering fee discounts for future interactions within the protocol to foster sustained user engagement. Multi-round airdrops can also enable protocols to refine strategies and address previously identified issues. Furthermore, developing more nuanced metrics for airdrop eligibility can mitigate airdrop farming by targeting genuine users rather than opportunistic actors. However, insider trading risks may persist under such frameworks.

\paraib{Target Reputable Entities} Protocols should prioritize distributing tokens to established entities, such as reputable projects, research groups, tech communities, and students. Funding these entities or allocating airdrops to DAOs developing relevant dApps can foster long-term involvement and align incentives with the protocol’s mission.

\paraib{Proactive Oversight and Community Engagement} Conducting thorough analyses of past airdrops is key, even those on different chains, to identify potential pitfalls and refine strategies. Maintaining open communication channels with the community can further align protocol developments with user interests.

\paraib{Align Rewards with Costs} Ensuring that rewards scale proportionately with associated costs can help deter airdrop farmers from exploiting the system. Aligning incentives can also foster a more equitable distribution of tokens to genuinely engaged participants.

In conclusion, these guidelines provide actionable strategies for blockchain protocols, researchers, and practitioners seeking to design more effective airdrop campaigns. Additionally, we release the dataset and scripts used in this study to encourage further research and reproducibility in the field~\cite{Messias-DataSet-Code-2025}. Furthermore, we believe that exploring the efficacy of ``lockdrops,'' where tokens are distributed but remain locked for a period, can potentially mitigate rapid sell-offs and promote long-term engagement. We leave this to future work.

%------------------------------------------------------------------------------

\bibliographystyle{plainurl}% the mandatory bibstyle
\bibliography{references}

\appendix

\section{Gemstone: A Case of Sybil Farming on ZKsync Era}\label{sec:gemstone}

Between July 23\tsup{rd} and July 31\tsup{st}, 2023, a Sybil farming campaign was carried out on ZKsync Era under the pretense of a token airdrop, known as Gemstone~\cite{Gemstone,Gemstone-Twitter}. Unlike legitimate airdrops that aim to incentivize long-term user engagement, Gemstone's campaign was explicitly designed to exploit anticipated eligibility criteria for a future ZKsync airdrop.
The orchestrator employed more than \num{20000} Sybil accounts and deployed a closed-source \gls{DEX} to simulate user activity. Although marketed as a token airdrop, the Gemstone token had no intrinsic value and remains worthless.

\begin{figure*}[t]
  \centering
  \includegraphics[width=\onecolgrid]{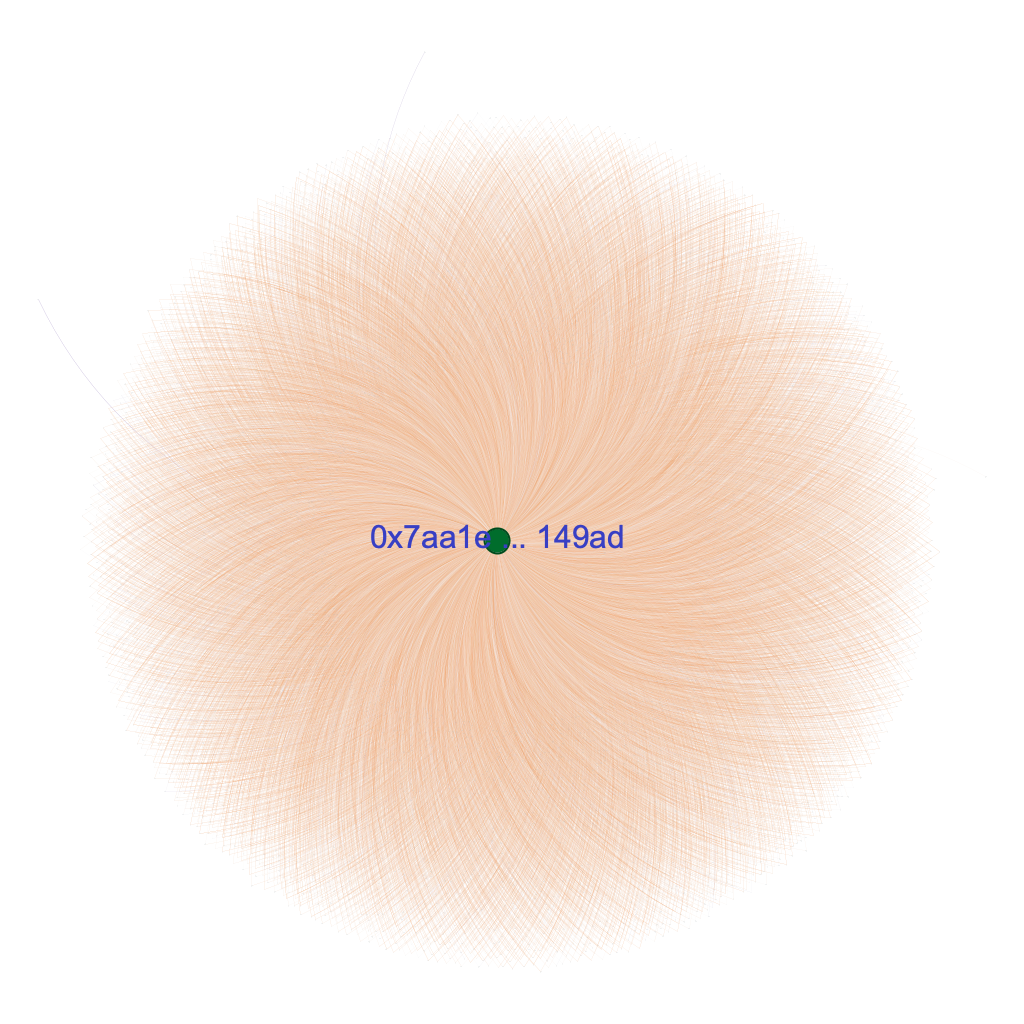}
  \caption{
    Network graph of the Gemstone airdrop, illustrating the structure of Sybil accounts transferring tokens to a self-developed, closed-source exchange. The figure highlights the centralization and coordination of token flows typical of Sybil farming behavior.
  }
  \label{fig:airdrop-gemstone-network}
\end{figure*}

Gemstone stands out due to the scale and structure of its distribution. Approximately 99.53\% of the token supply was distributed during the short campaign period—an outlier compared to all other airdrops we examined (as discussed in \S\ref{sec:analysis}). The airdrop also exhibited an unusually high median number of tokens per recipient, reflecting a heavily concentrated distribution pattern.
Notably, per Figure~\ref{fig:airdrop-gemstone-network}, 95\% of the distributed tokens were routed through fake exchanges created by the farmer, with nearly all ultimately consolidated into a single address (\href{https://era.zksync.network/address/0x7aa1ed8fb5e820e38c86cf3dd0a9eb0169a149ad#tokentxns}{0x7aa$\cdots$49ad}) via a single transaction hop. This rapid and centralized token flow indicates highly coordinated activity, consistent with a Sybil attack rather than legitimate user participation.

\section{More Insights Into Airdrop Metrics} \label{sec:airdrop_metrics}

\begin{figure}[t]
    \centering
    \includegraphics[width=\onecolgrid]{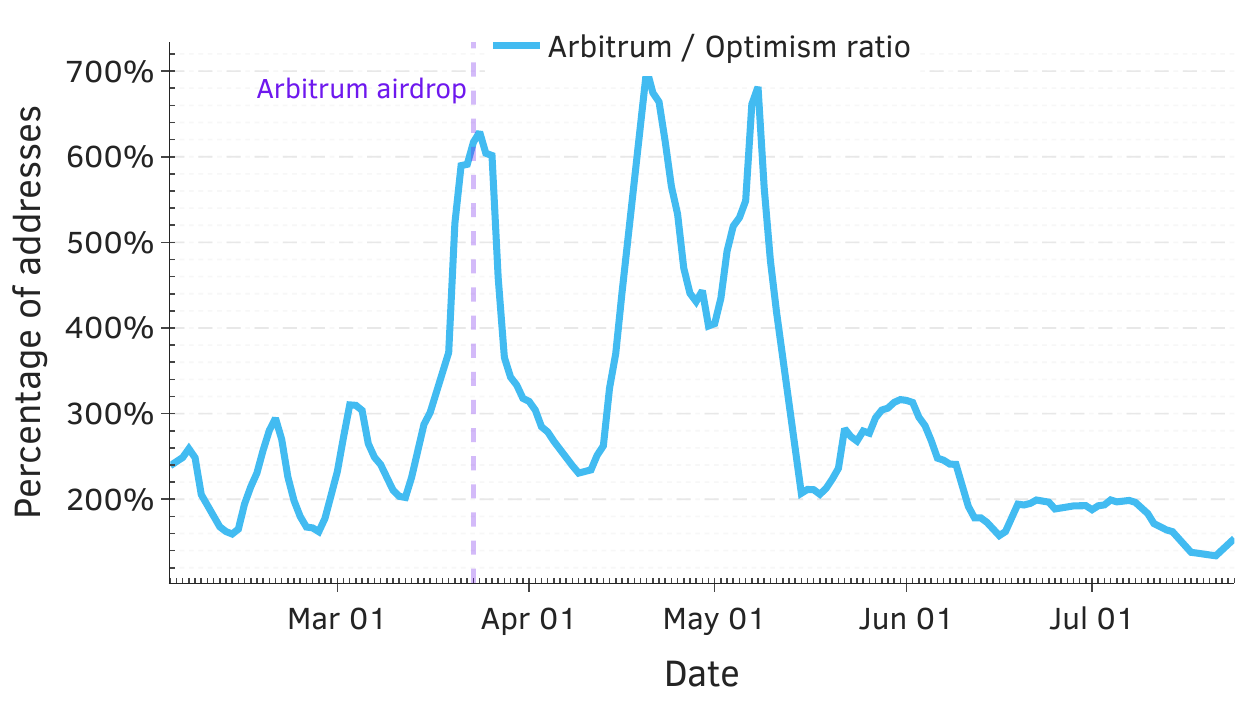}
    \caption{Ratio between the number of unique daily active addresses in Arbitrum and Optimism.}
    \label{fig:DailyActiveAddressesRatio}
\end{figure}

\begin{figure}[t]
    \centering
    \begin{subfigure}{\twocolgrid}    
        \centering
    \includegraphics[width=\twocolgrid]{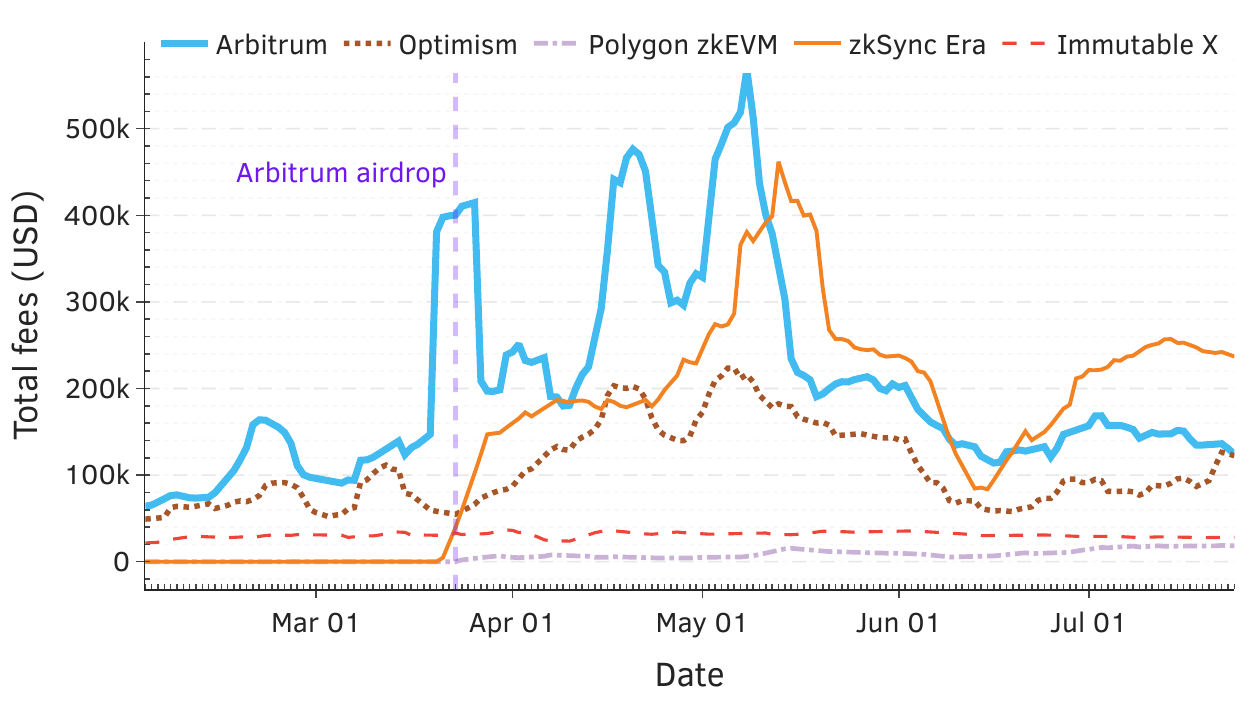}
        \caption{Absolute daily transaction fees.}
    \end{subfigure}
    \begin{subfigure}{\twocolgrid}
    \includegraphics[width=\twocolgrid]{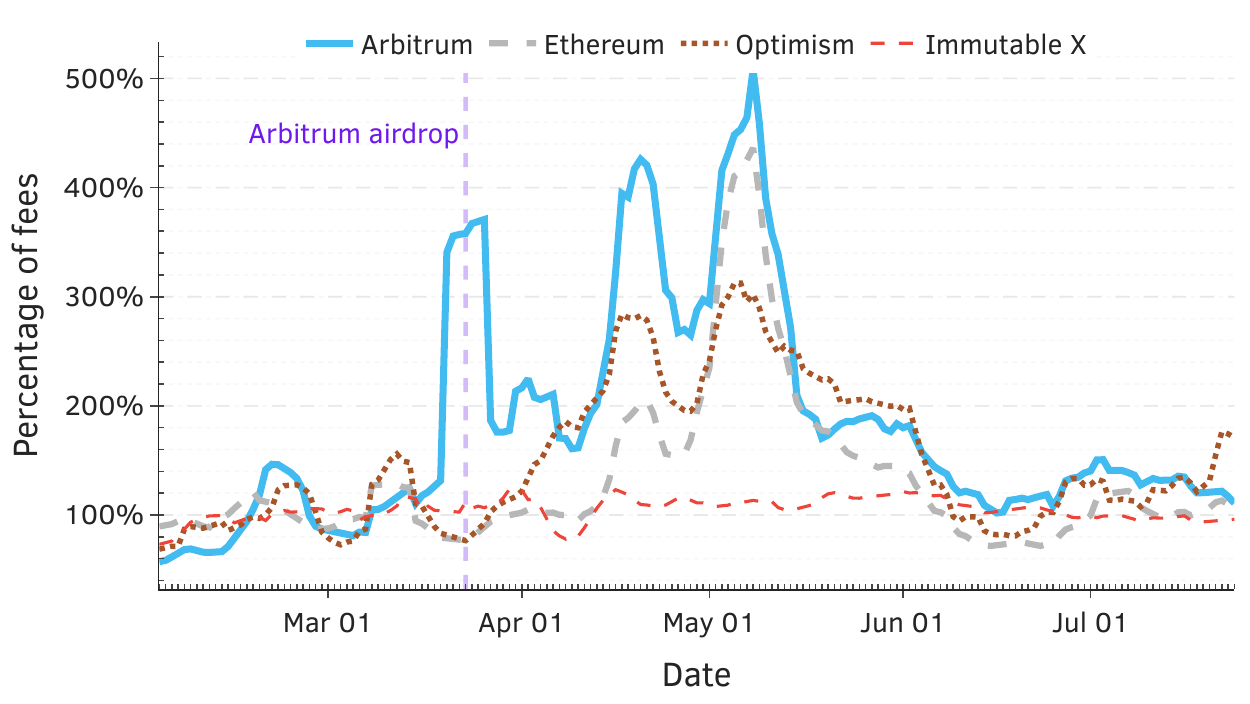}
        \caption{Relative to average.}
    \end{subfigure}
    \caption{Total daily transaction fees before and after the Arbitrum airdrop: (a) absolute values of daily fees in USD; and (b) fees relative to the average before the airdrop.}
    \label{fig:fees-before-after}
\end{figure}

In this section, we provide a complementary analysis of the measures described in Section \S\ref{sec:status-quo}. To summarize, Figure~\ref{fig:DailyActiveAddressesRatio} shows the ratio of unique daily active addresses between Arbitrum and Optimism. Additionally, Figure~\ref{fig:fees-before-after} presents the total daily transaction fees before and after the Arbitrum airdrop.

\section{Glossary}
\label{sec:glossary}
Following is a list of important notations used in this paper.
% START Make sure that the symbols and acronyms have constant indentation
% \renewcommand{\glossarysection}[2][]{}  % Disables the titles
\setglossarystyle{alttree}
\glssetwidest{AAAA}
% END
% \printnoidxglossary[type={symbols}]
\printnoidxglossary[type={acronym}]

\end{document}